\documentclass[aps,prl,reprint,10pt,superscriptaddress,longbibliography]{revtex4-2}

\usepackage[T1]{fontenc}
\usepackage[utf8]{inputenc}
\usepackage{amsmath,amssymb,bm,mathtools}
\usepackage{graphicx}
\usepackage{xcolor}
\usepackage{booktabs}
\usepackage{hyperref}
\hypersetup{colorlinks=true,urlcolor=[rgb]{0,0,0.36},linkcolor=[rgb]{0,0,0.36},citecolor=[rgb]{0,0,0.36}}
\usepackage[mathlines]{lineno}
\usepackage[normalem]{ulem}

\newcommand{\ETH}{Institute for Theoretical Physics, ETH Zurich, Wolfgang Pauli Strasse 27, 8093 Zurich, Switzerland}
\newcommand{\PKS}{Max Planck Institute for the Physics of Complex Systems, N\"{o}thnitzer Strasse 38, 01187 Dresden, Germany}
\newcommand{\Harvard}{Lyman Laboratory, Department of Physics, Harvard University, Cambridge, MA 02138, USA}
\newcommand{\Caltech}{Department of Physics and Institute for Quantum Information and Matter, California Institute of Technology, Pasadena, California 91125, USA}
\newcommand{\MIT}{Department of Physics, MIT, 77 Massachusetts Avenue, 02139 Cambridge, MA, USA}
\newcommand{\MPSD}{Max Planck Institute for the Structure and Dynamics of Matter, Luruper Chaussee 149, 22761 Hamburg, Germany}
\newcommand{\Oxford}{Clarendon Laboratory, University of Oxford, Parks Road, Oxford OX1 3PU, UK}
\usepackage{verbatim}
\newcommand{\avg}[1]{\left\langle #1\right\rangle}
\newcommand{\dd}{\mathrm{d}}
\newcommand{\ii}{\mathrm{i}}

\newcommand{\vecq}[1]{\vec{#1}}

\newcommand{\draftgraphic}[3]{%
  \IfFileExists{#1}{\includegraphics[width=#2]{#1}}{%
  \fbox{\parbox[c][0.22\textheight][c]{0.95\columnwidth}{\centering\textbf{Figure placeholder}\\[4pt]#3\\[4pt]\texttt{\detokenize{#1}}}}}}

\begin{document}

\title{Optical and magnetic signatures of drive-enhanced\\coherence in phase-disordered superconducting bilayers}

\author{Duilio De Santis}
\thanks{Contact author: ddesantis@phys.ethz.ch}
\affiliation{\ETH}
\author{Sambuddha Chattopadhyay}
\affiliation{\Harvard}
\affiliation{\ETH}
\author{Marios H. Michael}
\affiliation{\PKS}
\author{Andrea Cavalleri}
\affiliation{\MPSD}
\affiliation{\Oxford}
\author{Gil Refael}
\affiliation{\Caltech}
\author{Patrick A. Lee}
\affiliation{\MIT}
\author{Eugene A. Demler}
\affiliation{\ETH}

\date{\today}

\begin{abstract}
Recent experiments demonstrating superconducting-like characteristics in 
materials above their equilibrium transition temperatures following strong 
sub-picosecond optical excitation have stimulated theoretical interest in 
phase-disordered superconductors subject to a periodically modulated superfluid stiffness. 
Previous analyses of the single-layer optical conductivity have shown that such a 
drive can enhance the low-frequency $1/\omega$ response, even when the period-averaged 
stiffness remains unchanged. We extend these studies to bilayer systems, such as the YBa$_2$Cu$_3$O$_{6+x}$
(YBCO) cuprate, by considering drives with different symmetries. Specifically, we investigate 
a layer-symmetric drive, in which the two planes stiffen and soften simultaneously, 
and a layer-antisymmetric drive, in which one plane stiffens while the other softens.
Both drives enhance correlations of the layer-averaged phase and generate comparable 
superconducting-like optical conductivities. Only the symmetric drive, however, strengthens 
relative-phase locking and the associated counterflow response to a magnetic field 
parallel to the planes. Thus, an enhanced $1/\omega$ optical conductivity does not by itself 
imply enhanced magnetic screening. In view of applications to driven 
YBCO, where superconducting-like in-plane optics and transient diamagnetism 
have both been reported, our result provides a symmetry diagnostic: 
within a phase-synchronization mechanism, the experimentally active low-frequency 
stiffness modulation must contain a substantial layer-symmetric component.
\end{abstract}

\maketitle

\textit{Introduction.---} Symmetry-resolved measurements are a ubiquitous 
tool for elucidating microscopic mechanisms underlying photo-induced phenomena 
in driven quantum materials~\cite{delaTorre2021Nonthermal}. For example, polarization-resolved 
transient reflectivity was used to separate symmetric and nematic superconducting collective 
responses in Ba$_{1-x}$K$_x$Fe$_2$As$_2$ ~\cite{Grasset2022NematicMode}; time-resolved 
reflection anisotropy used to extricate the melting of C$_4$-breaking charge-orbital order 
in a layered manganite~\cite{PerezSalinas2022Multimode}; time-resolved second-harmonic 
generation used to separately track the recovery of anti-ferromagnetic order from photo-induced 
charge dynamics in driven Sr$_2$IrO$_4$~\cite{delaTorre2022DrivenMott}. The interrogative 
power of these measurements arises from the fact that different 
coherently driven collective modes transiently modify particular channels of the 
material response tensor, leaving their traces when those specific channels 
are probed upon driving. 

Such symmetry aspects have received comparatively little attention in 
the interpretation of light-induced superconducting-like behavior~\cite{Fausti2011LightInduced,
Hu2014OpticallyEnhanced,Kaiser2014Optically,Mittrano2016Possible,Cavalleri2018Photo,Budden2021Evidence,
Buzzi2020Photomolecular,Liu2020PumpFrequencyResonances,Fava2024Magnetic,Rosenberg2025Signatures}. 
Experiments have reported both a low-frequency inductive optical 
response~\cite{Fausti2011LightInduced,Hu2014OpticallyEnhanced,Mittrano2016Possible,
Buzzi2020Photomolecular,Rosenberg2025Signatures} and transient diamagnetic screening 
far above the equilibrium superconducting transition~\cite{Fava2024Magnetic}. 
In a recent paper, several of us have offered an alternative explanation based on 
the production of a paramagnetic magnetization pulse due to an instability created by the drive~\cite{Michael2026FluxFloquet}. 
Here, we instead consider a distinct mechanism in which pump-induced modulation of 
the superfluid stiffness drives a transient enhancement of the diamagnetic response.
In equilibrium, inductive optical response and diamagnetic screening are 
complementary manifestations of phase rigidity. Out of equilibrium, however, they 
need not be equivalent: the two probes may couple different collective modes, and their 
coexistence may therefore contain information about the microscopic origin of the drive. 
A superconducting bilayer, the setting of several of the experiments
reporting light-induced superconducting-like signatures, provides a 
particularly direct setting in which to expose this distinction. In a bilayer, a spatially 
uniform in-plane probe electric field produces parallel currents in the two layers and the optical 
conductivity primarily provides information on the correlations of the \textit{layer-averaged} phase. By contrast, 
a magnetic field parallel to the planes generates oppositely directed currents: diamagnetic 
screening arises from long-range coherence in the \textit{layer-relative} phase, 
or counterflow, sector~\cite{Zhang2005Vortices, Homann2024DissipationlessCounterflow,
Michael2026Counterflow}. This raises a general question: can distinct driving 
mechanisms produce qualitatively similar superconducting-like optical responses 
while generating different magnetic responses?

\begin{figure}[t]
\centering
\draftgraphic{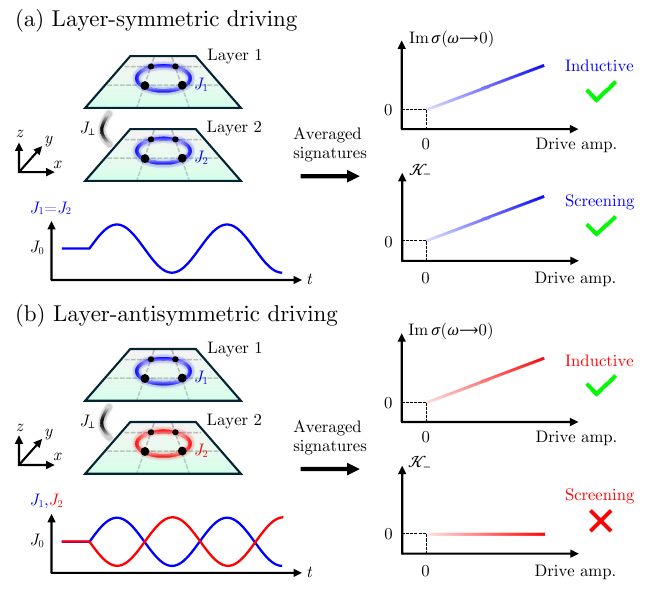}{\columnwidth}{Driven bilayer schematic}
\caption{\textbf{Driven bilayer schematics.}
Periodic stiffness modulation can produce both an enhanced low-frequency
inductive response and enhanced magnetic screening in a driven,
phase-disordered bilayer. For a layer-symmetric drive [panel~(a)], the two
planes stiffen and soften together, enhancing long-distance correlations in
both the average ($+$) and relative ($-$) phase sectors. For a
layer-antisymmetric drive [panel~(b)], one plane stiffens while the other
softens, so that average-phase correlations can still be enhanced, whereas
relative-phase correlations are not. A uniform in-plane optical probe couples
to the parallel ($+$) current and measures the conductivity $\sigma$,
whereas a magnetic field parallel to the layers probes the counterflow ($-$)
screening channel characterized by $\mathcal K_-$; these responses are defined formally
below. Consequently, both drives can enhance the $1/\omega$ optical response,
but only the symmetric drive simultaneously enhances magnetic screening.}
\label{fig:schematic}
\end{figure}

Here we answer this in the affirmative. We establish that 
dynamical suppression of phase fluctuations~\cite{DeSantis2025Enhanced, Diessel2026SteadyStates} 
can enhance both the in-plane superconducting-like optical response and magnetic 
screening in a driven bilayer, without invoking an increase in the cycle-averaged 
stiffness, consistent with experiments~\cite{Rosenberg2025Signatures, Fava2024Magnetic}. 
Crucially, however, a strong $1/\omega$ optical response can appear without enhanced 
magnetic screening: whether the two signatures occur together is controlled by the 
layer symmetry of the stiffness modulation. In the language of the Lawrence–Doniach 
description of superconducting phase dynamics introduced below, we write the layer 
stiffness as $J_\ell(t)=J_0+\Delta J_\ell(t)$ ($\ell=1,2$ is the layer index) 
and compare a layer-symmetric drive, $\Delta J_1=\Delta J_2$, with a 
layer-antisymmetric drive, $\Delta J_1=-\Delta J_2$, see Fig.~\ref{fig:schematic}. 
Both channels are, in principle, accessible in YBa$_2$Cu$_3$O$_{6+x}$ (YBCO) 
through driven phonon dynamics~\cite{Mankowsky2014Nonlinear,Fechner2016EffectsOfIntense}. Starting from 
a metallic initial state above the phase-disordering temperature, both drives 
enhance layer-averaged phase correlations and generate a pronounced $1/\omega$ 
optical response, see Fig.~\ref{fig:schematic}(a)-(b). The antisymmetric drive, 
however, leaves relative-phase correlations and magnetic screening essentially 
unchanged, see Fig.~\ref{fig:schematic}(b). Within our driven phase-coherence 
enhancement mechanism~\cite{DeSantis2025Enhanced}, the simultaneous 
observation of superconducting-like optical and magnetic signatures necessitates 
a substantial layer-symmetric component to the drive. Combining optical and 
magnetic probes therefore provides significant discriminatory insight into 
possible microscopic mechanisms for light-induced superconducting-like
behavior.

\textit{Driven bilayer model.---}
We consider a bilayer XY model inspired by the pseudogap physics of the
bilayer cuprate YBCO in the underdoped regime, where local superconducting correlations may persist
above $T_c$ despite strong phase fluctuations~\cite{Dubroka2011Evidence,Uykur2014Persistence}.
Within a Lawrence--Doniach description%
~\cite{Lawrence1971LayerStructureSuperconductors,Bulaevskii1992Vortices},
the dimensionless Hamiltonian is
(see~\footnote{A superscript $*$ denotes a dimensionless quantity.  We use
$H^{*}=H/J_0$, $J_\ell^{*}=J_\ell/J_0$,
$J_\perp^{*}=J_\perp/J_0$, $T^{*}=k_{\rm B}T/J_0$,
$t^{*}=t/\tau=\Gamma J_0t$, and $\Omega^{*}=\Omega\tau$, with
$\tau=(\Gamma J_0)^{-1}$. Here $\Gamma$ is the kinetic 
coefficient that enters in the model-A dynamical equation.  The Peierls phase
$A_{\ell,ij}^{*}=(2e/\hbar)\int_i^j\mathbf A_\ell\!\cdot\dd\boldsymbol\ell$
is dimensionless as well.  To lighten the notation, we drop the superscript
$*$ throughout the remainder of the main text.} for notational details)
\begin{equation}
\begin{aligned}
H
={}&-\sum_{\ell=1}^{2}J_\ell
\sum_{\langle ij\rangle}
\cos\!\left(
\theta_{\ell i}-\theta_{\ell j}-A_{\ell,ij}
\right)\\
&-J_\perp\sum_i
\cos\!\left(\theta_{1i}-\theta_{2i}\right).
\end{aligned}
\label{eq:H}
\end{equation}%
Here $\theta_{\ell i}$ is the superconducting phase at site $i$
of layer $\ell=1,2$. In equilibrium, $J_\ell=1$, while $J_\perp\ll1$ is
the dimensionless intrabilayer Josephson coupling~\footnote{
For an anisotropic phase-only model, the ratio of interlayer to in-plane
couplings can be estimated from the zero-temperature penetration depths.
We write~\cite{Mihlin2009TemperatureDependence}
$
J_\perp
\sim
\left(
\frac{\lambda_{ab} \; a}{\lambda_c \; d_\perp}
\right)^2,
$
where $a\simeq3.8\,$\AA{} is the in-plane lattice constant and
$d_\perp\simeq4\,$\AA{} is the intrabilayer separation. Using the YBCO
penetration-depth values quoted in
Ref.~\cite{Mihlin2009TemperatureDependence} gives an intrabilayer coupling
ratio of order
$
J_\perp\sim10^{-3}\text{--}10^{-2}.
$
We therefore use $J_\perp\sim0.01$ as a representative weak-coupling value
near the upper end of this experimentally motivated range.}.  The Peierls phase
$A_{\ell,ij}$ accounts for coupling to an electromagnetic vector potential.
This phase-only description neglects amplitude fluctuations of the
superconducting order parameter but retains thermal phase fluctuations and
vortices.

Motivated by the phonon dynamics induced by intense THz and mid-infrared
irradiation, we allow the in-plane stiffness of each layer to acquire a
time-dependent modulation, $J_\ell(t)=1+\Delta J_\ell(t)$. We consider a
layer-symmetric modulation, $\Delta J_1(t)=\Delta J_2(t)$, and a
layer-antisymmetric modulation, $\Delta J_1(t)=-\Delta J_2(t)$. We
parameterize the two cases as
$\Delta J_1(t)=\pm\Delta J_2(t)=f(t)\,A\sin(\Omega t)$, where $f(t)$
is a smooth Gaussian-like switch-on, $A$ is the stiffness-modulation 
amplitude, and $\Omega$ is the angular drive frequency.  
We refer the reader to the Appendix of Ref.~\cite{DeSantis2025Enhanced} 
for estimates and a discussion of possible microscopic mechanisms 
underlying such modulations in driven cuprates. Here, we consider 
symmetric (Raman-like) and antisymmetric (IR-like) modulations on 
equal footing, in order to theoretically isolate the consequences 
of different drive symmetries for the bilayer response.

As appropriate for experiments performed above the critical 
temperature of high-$T_c$ superconductors such as YBCO, 
we consider overdamped model-A dynamics for the phase degrees 
of freedom~\cite{Hohenberg1977Theory,Podolsky2007Nernst},
\begin{equation}
\begin{aligned}
\partial_t\theta_{\ell i}
&=-\frac{\partial H}{\partial\theta_{\ell i}}
+\xi_{\ell i}(t),\\
\avg{\xi_{\ell i}(t)\xi_{\ell'j}(t')}
&=2T\,\delta_{\ell\ell'}\delta_{ij}\delta(t-t').
\end{aligned}
\label{eq:modelA}
\end{equation}
Here $\xi_{\ell i}$ is a Gaussian thermal noise source. 
The natural bilayer variables are the average and relative phases,
\begin{equation}
\theta_+(\vecq r)
=\frac{\theta_1(\vecq r)+\theta_2(\vecq r)}{2},
\qquad
\theta_-(\vecq r)
=\theta_1(\vecq r)-\theta_2(\vecq r).
\label{eq:pmphase}
\end{equation}
In the simulations, we implement the compact layer phases $\theta_1$ and
$\theta_2$, while $\theta_+$ and $\theta_-$ are constructed from them using
the appropriate branch-cut convention, as described in the Supplemental Material~\cite{Supplement}.
Numerical details, including the integration scheme, drive switch-on protocol, and
convergence tests, are also given there.

\textit{Driven $G_\pm(r)$ correlations.---}
We monitor the equal-time, radially averaged phase correlators
\begin{equation}
G_\pm(r)
=\overline{
\avg{
\cos\!\left[
\theta_\pm(\vecq r_0+\vecq r,t)
-\theta_\pm(\vecq r_0,t)
\right]
}},
\label{eq:corr}
\end{equation}
where $\avg{\cdots}$ denotes spatial and noise averaging, and the overbar
denotes a time average in the periodically driven state.  The correlator $G_+$ 
characterizes correlations of the layer-averaged phase, whereas $G_-$ characterizes 
relative-phase correlations and hence the degree of interlayer phase locking. As we 
discuss further below, these two sectors are respectively associated with the 
optical-conductivity response and the counterflow Meissner response to 
a magnetic field parallel to the layers.

\begin{figure}[t]
\centering
\draftgraphic{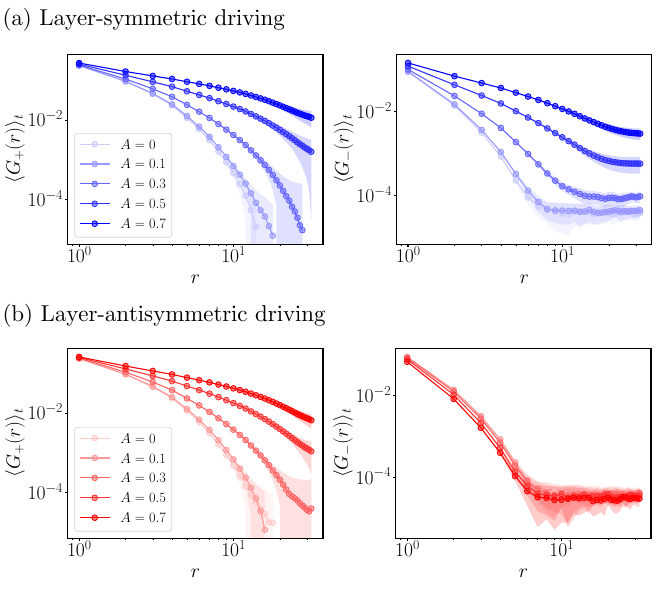}{\columnwidth}{Correlations}
\caption{\textbf{Average and relative phase correlations under layer-symmetric and layer-antisymmetric modulation.}  
Time-averaged correlations as a function of distance, for increasing values of $A$ (see legends in 
the left panels), with $T=1.5$, $\Omega=0.01$, $J_\perp=0.01$, $L=64$, and $N=50$ independent realizations.  
Symmetric driving enhances both $G_+$ and $G_-$, see panel~(a).  Antisymmetric driving produces a 
comparable enhancement of $G_+$ but little or no enhancement of $G_-$, see panel~(b).  
Shaded areas denote the statistical error over realizations.}
\label{fig:corr}
\end{figure}

Figure~\ref{fig:corr} summarizes a key dynamical result.  At the chosen temperature above the 
equilibrium ordering crossover (we discuss equilibrium benchmarks in the Supplemental Material~\cite{Supplement}), 
the undriven bilayer has short-ranged $G_+$ and a finite but modest relative-phase correlation 
set by $J_\perp$~\cite{Fertig2002Deconfinement,Zhang2005Vortices,Zhang2006Correlation,Michael2026Counterflow}, 
see the $A=0$ curves in Fig.~\ref{fig:corr}.  
Increasing the amplitude of a slow symmetric drive enhances large-distance correlations in both channels, 
see Fig.~\ref{fig:corr}(a).  This is a bilayer analogue of the drive-enhanced coherence found in a 
single layer~\cite{DeSantis2025Enhanced}: during the high-stiffness portion of the cycle, vortices annihilate 
and phase correlations grow, whereas the subsequent low-stiffness interval does not fully undo this gain because 
of the nonlinear character of the system's dynamics. The frequency- and amplitude-dependence is similar to that found 
in the single-layer case and is presented for completeness in the Supplemental Material~\cite{Supplement}.

Under the same equilibrium parameters and drive frequency and amplitude, the response to layer-antisymmetric 
driving is qualitatively different, see Fig.~\ref{fig:corr}(b).  Notably, its large-distance $G_+$ enhancement can be as large as in the 
symmetric case, despite the absence of any modulation of the instantaneous layer average $(J_1+J_2)/2=1$.  
The relative channel, however, does not share this gain: $G_-(r)$ remains close to equilibrium and can decrease at strong driving.  
This conclusion is robust across the slow-drive region of the frequency--amplitude plane~\cite{Supplement}.  
A possible interpretation is that the presence of a soft layer at every instant disfavors interlayer locking, 
while the weak $J_\perp$ is insufficient to transfer the gain in in-plane coherence between 
the layers on the timescale of the drive.

\textit{Optical conductivity.---}
A spatially uniform in-plane electric field (say, along the $x$ direction)
produces the same Peierls phase in the two layers,
$A_{1x}=A_{2x}=A_p$. It therefore couples to the symmetric dimensionless
current
\begin{equation}
\begin{aligned}
j_+=\tfrac12(j_1+j_2), \; \mathrm{where} \; j_\ell =\frac{1}{L^2}\sum_i
J_\ell(t)\sin(\Delta_x\theta_{\ell i}-A_p) .
\end{aligned}
\label{eq:jplus}
\end{equation}
In both the equilibrium and driven scenarios, we apply a weak monochromatic
probe $E_p(t)=E_0\sin(\omega t)$, with $E_p=-\partial_tA_p$.  The
frequency-resolved dimensionless conductivity is
\begin{equation}
\sigma(\omega)=\frac{j_+(\omega)}{E_p(\omega)}.
\label{eq:sigma}
\end{equation}
Each probe frequency is obtained from an independent simulation. In
extracting the conductivity in Eq.~\eqref{eq:sigma}, we retain only the
Fourier component at the probe frequency $\omega$ and exclude the Floquet
sidebands at $\omega\pm m\Omega$, as is customary when analyzing the
pump-modified optical response in ultrafast experiments%
~\cite{Cavalleri2018Photo,Liu2020PumpFrequencyResonances,
Rosenberg2025Signatures}. Linearity of the response in the dimensionless probe field
$E_0$ is explicitly verified in the Supplemental Material~\cite{Supplement}.

\begin{figure}[t]
\centering
\draftgraphic{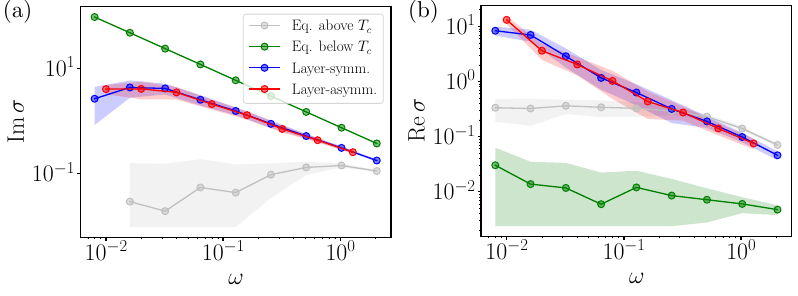}{\columnwidth}{Conductivity}
\caption{\textbf{Optical response under layer-symmetric and layer-antisymmetric driving.}
Imaginary [panel~(a)] and real [panel~(b)] parts of the conductivity $\sigma$ in
equilibrium and under layer-symmetric and layer-antisymmetric 
driving, evaluated under the favourable conditions identified in
Fig.~\ref{fig:corr}. For both drive symmetries, the dissipative response
narrows and the low-frequency $1/\omega$ behavior of
$\operatorname{Im}\sigma$ is enhanced, as seen by comparing the driven
curves (blue: symmetric, red: antisymmetric, both with $A=0.7$) 
with the equilibrium curves (gray, $A=0$). The close
similarity between the two driven responses reflects the comparable enhancement of
$G_+$. For reference, we also show the equilibrium superconducting-like response below
the phase-disordering transition (in green), characterized by a $1/\omega$ divergence in
the imaginary part and a negligible real part. Shaded areas denote the statistical 
error over realizations. Other parameters are
$L=64$, $N=50$, $T=1.5$ for the driven and equilibrium curves,
$T=0.7$ for the low-temperature reference, $J_\perp=0.01$,
$\Omega=0.01$, and dimensionless probe amplitude ranging $E_0=0.0005$--$0.001$.}
\label{fig:conductivity}
\end{figure}

For equilibrium parameters and drive frequency and amplitude comparable to
those used in Fig.~\ref{fig:corr}, Fig.~\ref{fig:conductivity} shows that
the optical conductivity does not distinguish between the two drive
symmetries. Both produce a narrower low-frequency dissipative feature and
an enhanced imaginary response approaching $D_{\rm eff}/\omega$. This
behavior tracks the enhancement of $G_+$ and resembles the
phenomenology found in the driven single-layer model%
~\cite{DeSantis2025Enhanced}. In particular, the antisymmetric drive 
shows that a pronounced $1/\omega$-like optical response can 
emerge even when the instantaneous layer-averaged stiffness 
remains strictly constant, $(J_1+J_2)/2=1$.

\textit{Magnetic screening and counterflow.---}
We next probe the counterflow channel by applying opposite static Peierls 
phases to the two layers, $A_{1x}=-A_{2x}=\alpha_B$. This configuration corresponds 
to a static magnetic field parallel to the layers, which we take as $\vecq B=B_y\vecq e_y$. 
In the gauge $A_x(z)=B_y z$, the two layers, located at $z=\pm d_\perp/2$, 
acquire equal and opposite Peierls phases. The resulting response is a counterflow current,
\begin{equation}
\begin{aligned}
j_-=\tfrac12(j_1-j_2), \; \; \mathcal K_-
=-\lim_{\alpha_B\to0}
\frac{\overline{\avg{j_-}}}{\alpha_B}.
\end{aligned}
\label{eq:Kminus}
\end{equation}
The response is linear over the field range used in the simulations~\cite{Supplement}.  
Unlike a uniform optical probe, this antisymmetric gauge perturbation couples directly to $\theta_-$: 
in a gauge with $A_x=0$, the same flux appears as a spatially varying phase in the 
interlayer Josephson term, making the relative-phase origin of the response explicit.
For completeness, because we impose open boundary conditions along 
the $x$ direction for this calculation, we can directly resolve the full spatial profile of the 
screening currents, as detailed in the Supplemental Material~\cite{Supplement}.

\begin{figure}[t]
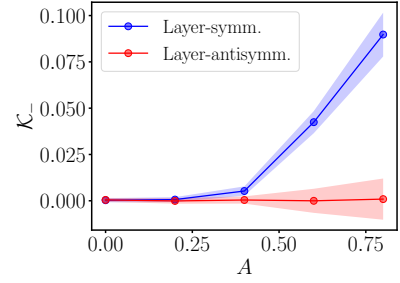

\centering
\draftgraphic{fig4_1col.pdf}{0.6\columnwidth}{Screening}
\caption{\textbf{Counterflow screening response under layer-symmetric and
layer-antisymmetric driving.}
{Counterflow screening kernel evaluated under driving conditions similar to those in
Fig.~\ref{fig:conductivity}. The symmetric drive (in blue) substantially
enhances the response to a static magnetic field parallel to the planes,
bringing it to a significant fraction of the corresponding equilibrium
response below the phase-disordering temperature, shown for reference in the
Supplement~\cite{Supplement}. By contrast, under antisymmetric driving (in red), the response
remains close to its equilibrium value. This behavior mirrors that of $G_-$
in Fig.~\ref{fig:corr}. Shaded areas denote the statistical error over realizations.
Other parameters are $L=64$, $N=50$, $T=1.5$, $J_\perp=0.02$, $\Omega=0.005$, and $\alpha_B=0.04$.}
}
\label{fig:screening}
\end{figure}

Under symmetric driving, $\mathcal K_-$ increases together with $G_-$, see 
Fig.~\ref{fig:screening} (blue curve). Thus, this driven state exhibits both an
enhanced inductive optical response and enhanced magnetic screening. By
contrast, antisymmetric driving produces a comparable enhancement of
$\operatorname{Im}\sigma$ but no corresponding increase in magnetic
screening, see Fig.~\ref{fig:screening} (red curve). The striking contrast revealed by
comparing Figs.~\ref{fig:conductivity} and~\ref{fig:screening} shows
that, out of equilibrium, the $1 / \omega$ enhancement of the optical 
conductivity does not imply a correspondingly enhanced magnetic screening 
response, underscoring the necessity of these complementary 
probes of light-induced superconducting-like behavior.

Magnetic screening is a counterflow response and therefore requires coherent current flow in both
layers. As antisymmetric driving leaves one layer in the low-stiffness
part of the cycle at any given time, the weaker layer limits the
counterflow response and prevents the gain in in-plane coherence from
developing into enhanced interlayer magnetic screening. A spatially uniform 
in-plane optical field, by contrast, probes the parallel-flow channel, 
which primarily tracks the layer-averaged phase.

This distinction is conceptually relevant to the interpretation of the YBCO
experiments. A divergent in-plane imaginary conductivity has been reported
following mid-infrared excitation of apical-oxygen modes%
~\cite{Rosenberg2025Signatures}, while transient diamagnetism has been
observed under closely related driving conditions%
~\cite{Fava2024Magnetic}. If both responses arise from drive-enhanced
phase coherence of preformed pairs, our results rule out an 
antisymmetric low-frequency modulation of the in-plane stiffness. The
effective modulation must instead contain a substantial layer-symmetric
component. This conclusion follows from the fundamental symmetry structure 
of the driven bilayer response, rather than from microscopic details.

\textit{Remarks on applicability to driven YBCO.---}
The present work significantly advances the phase-synchronization scenario of
Ref.~\cite{DeSantis2025Enhanced} by incorporating the bilayer structure
characteristic of YBCO.  Within a single framework that explicitly retains
thermal phase fluctuations and vortices, we find that periodic stiffness
modulation can reproduce, at a qualitative level, the two principal
superconducting-like signatures reported in driven YBCO: an enhanced
low-frequency inductive optical response and an enhanced screening response
to a magnetic field parallel to the planes.  The simultaneous appearance of
these effects is encouraging for a phase-ordering interpretation of the
experiments.  At the same time, quantitative comparison with YBCO raises
important aspects that lie beyond the minimal bilayer model considered here.

For the optical response, an intrinsic dynamical scale follows from
the mechanism responsible for the coherence enhancement.  As discussed in
Ref.~\cite{DeSantis2025Enhanced}, long-distance correlations develop
predominantly during the high-stiffness portion of each drive cycle.  Their
growth is limited by the slow dynamics of phase defects, giving a
drive-dependent growing length whose maximum value is controlled, up to
order-one factors, by the time available within a drive period.  Consequently,
the driven state does not acquire superconducting-like features on
arbitrarily long length and time scales.  This feature is visible 
in the conductivity: the drive narrows the dissipative feature in
$\operatorname{Re}\sigma$ and produces a corresponding $1/\omega$-like
regime in $\operatorname{Im}\sigma$, but the resulting effective scattering
scale remains tied to the drive frequency.  In Fig.~\ref{fig:conductivity},
for example, the crossover away from the enhanced inductive response occurs
at a frequency of the same order as $\Omega$.

This point becomes relevant when quantitatively comparing with the transient in-plane
optical response of YBCO.  Experimentally, the superconducting-like
$\operatorname{Im}\sigma\propto1/\omega$ behavior extends to probe
frequencies substantially below the characteristic frequency expected for
the effective stiffness modulation.  Even if the relevant low-frequency
drive is generated through nonlinear mixing of the approximately
$17$–$19 \, {\rm THz}$ apical-oxygen modes, giving a modulation at a frequency
of order a few THz, the measured inductive response extends to frequencies
of order $0.5 \, {\rm THz}$~\cite{Rosenberg2025Signatures}.  
This issue is distinct from the adiabaticity
condition emphasized in Ref.~\cite{DeSantis2025Enhanced}: a
difference-frequency drive can be sufficiently slow compared with the
microscopic relaxation rate to enhance correlations, while still being
too fast to account, within the present local conductivity calculation, for
an inductive response extending to much lower probe frequencies.  
Possible additional contributions may arise from Fresnel--Floquet
effects~\cite{Michael2022Generalized} in the experimentally inferred
conductivity, as well as from inertial phase dynamics and nonequilibrium
states away from the quasi-adiabatic limit. Incorporating these
effects into the connection between the driven response and the
experimentally inferred transient conductivity is an important direction
for future work.

A separate quantitative issue concerns magnetic screening. Here we have
isolated the intrabilayer physics and demonstrated that drive-enhanced
relative-phase coherence produces an enhanced counterflow response and,
consequently, an increased diamagnetic response to a magnetic field parallel
to the planes. The experiment, however, probes the magnetic response of a
macroscopic sample, for which screening currents generally involve
inter-bilayer current flow and electrodynamic coupling~\cite{Fava2024Magnetic}. 
Quantitatively accounting for the reported magnetic response is therefore 
likely to require physics beyond an isolated bilayer. A natural next step is 
to embed the present phase-fluctuating bilayer dynamics in a multi-bilayer electrodynamic
model coupled self-consistently to Maxwell equations. Such a treatment can
determine whether the drive-induced phase ordering found here produces a
macroscopic magnetic susceptibility of the magnitude observed
experimentally.

\textit{Conclusions.---}
Our results demonstrate that periodic suppression of phase fluctuations 
can generate both an enhanced low-frequency optical response and 
enhanced magnetic screening in a driven bilayer, providing a phase-only route 
to the simultaneous emergence of these superconducting-like signatures. A further central 
result is that these two responses are not equivalent and depend sharply 
on the symmetry of the drive.  The low-frequency $1/\omega$ optical response is mainly 
associated with long-distance correlations in the layer-average phase sector and can be 
enhanced by both layer-symmetric and layer-antisymmetric modulation.  Magnetic screening of a 
field parallel to the planes is a counterflow response and requires enhanced 
long-distance relative-phase correlations.  Only the layer-symmetric drive improves both.  
Thus the combination of optical conductivity and magnetic screening is a phase-sensitive 
probe of the symmetry of the microscopic drive.

The conclusion also clarifies the limits of optical evidence.  A $1/\omega$ contribution 
can arise from a long-lived collective current in the symmetric channel even when 
the system lacks a correspondingly enhanced counterflow stiffness.  Conversely, 
observing both a large optical inductive response and diamagnetism places a stronger 
constraint on any theory of the driven state.  In YBCO, it should contain a substantial 
modulation component that acts in phase on the two planes of each bilayer, consistent with nonlinear 
generation of a layer-symmetric Raman-type component.  More generally, symmetry-resolved probes 
of multilayer systems can distinguish genuine enhancement of the full superconducting 
phase rigidity from enhancement confined to a single transport channel.

\begin{acknowledgments}
We thank R. Andrei, U. Bhattacharya, M. Buzzi, J. Curtis, P.E. Dolgirev, M. Fechner, A. Gómez-Salvador, 
S.A. Kivelson, O. Malyshev, D. Nicoletti, M. Rosenberg, S. Sachdev, and A. Subedi for 
insightful discussions. D.D.S., S.C., and E.D. acknowledge support from 
SNSF Project No. 200021\_212899, SNSF Sinergia Grant No. CRSII--222792, 
and the Swiss State Secretariat for Education, Research and Innovation (contract No. UeM019-1).
G.R. acknowledges support from the AFOSR MURI program under 
agreement number FA9550-22-1-0339, as well as the Simons Foundation and the 
Aspen Center for Theoretical Physics. G.R. also acknowledges support 
from the Institute for Quantum Information and Matter. P.L. acknowledges 
support by DOE (USA) office of Basic Sciences Grant No. DE-FG02-03ER46076.
\end{acknowledgments}

%\bibliographystyle{apsrev4-2}
%\bibliography{refs_bilayer}

\begin{thebibliography}{38}%
\makeatletter
\providecommand \@ifxundefined [1]{%
 \@ifx{#1\undefined}
}%
\providecommand \@ifnum [1]{%
 \ifnum #1\expandafter \@firstoftwo
 \else \expandafter \@secondoftwo
 \fi
}%
\providecommand \@ifx [1]{%
 \ifx #1\expandafter \@firstoftwo
 \else \expandafter \@secondoftwo
 \fi
}%
\providecommand \natexlab [1]{#1}%
\providecommand \enquote  [1]{``#1''}%
\providecommand \bibnamefont  [1]{#1}%
\providecommand \bibfnamefont [1]{#1}%
\providecommand \citenamefont [1]{#1}%
\providecommand \href@noop [0]{\@secondoftwo}%
\providecommand \href [0]{\begingroup \@sanitize@url \@href}%
\providecommand \@href[1]{\@@startlink{#1}\@@href}%
\providecommand \@@href[1]{\endgroup#1\@@endlink}%
\providecommand \@sanitize@url [0]{\catcode `\\12\catcode `\$12\catcode
  `\&12\catcode `\#12\catcode `\^12\catcode `\_12\catcode `\%12\relax}%
\providecommand \@@startlink[1]{}%
\providecommand \@@endlink[0]{}%
\providecommand \url  [0]{\begingroup\@sanitize@url \@url }%
\providecommand \@url [1]{\endgroup\@href {#1}{\urlprefix }}%
\providecommand \urlprefix  [0]{URL }%
\providecommand \Eprint [0]{\href }%
\providecommand \doibase [0]{https://doi.org/}%
\providecommand \selectlanguage [0]{\@gobble}%
\providecommand \bibinfo  [0]{\@secondoftwo}%
\providecommand \bibfield  [0]{\@secondoftwo}%
\providecommand \translation [1]{[#1]}%
\providecommand \BibitemOpen [0]{}%
\providecommand \bibitemStop [0]{}%
\providecommand \bibitemNoStop [0]{.\EOS\space}%
\providecommand \EOS [0]{\spacefactor3000\relax}%
\providecommand \BibitemShut  [1]{\csname bibitem#1\endcsname}%
\let\auto@bib@innerbib\@empty
%</preamble>
\bibitem [{\citenamefont {de~la Torre}\ \emph {et~al.}(2021)\citenamefont
  {de~la Torre}, \citenamefont {Kennes}, \citenamefont {Claassen},
  \citenamefont {Gerber}, \citenamefont {McIver},\ and\ \citenamefont
  {Sentef}}]{delaTorre2021Nonthermal}%
  \BibitemOpen
  \bibfield  {author} {\bibinfo {author} {\bibfnamefont {A.}~\bibnamefont
  {de~la Torre}}, \bibinfo {author} {\bibfnamefont {D.~M.}\ \bibnamefont
  {Kennes}}, \bibinfo {author} {\bibfnamefont {M.}~\bibnamefont {Claassen}},
  \bibinfo {author} {\bibfnamefont {S.}~\bibnamefont {Gerber}}, \bibinfo
  {author} {\bibfnamefont {J.~W.}\ \bibnamefont {McIver}},\ and\ \bibinfo
  {author} {\bibfnamefont {M.~A.}\ \bibnamefont {Sentef}},\ }\href
  {https://doi.org/10.1103/RevModPhys.93.041002} {\bibfield  {journal}
  {\bibinfo  {journal} {Reviews of Modern Physics}\ }\textbf {\bibinfo {volume}
  {93}},\ \bibinfo {pages} {041002} (\bibinfo {year} {2021})},\ \bibinfo {note}
  {publisher: American Physical Society}\BibitemShut {NoStop}%
\bibitem [{\citenamefont {Grasset}\ \emph {et~al.}(2022)\citenamefont
  {Grasset}, \citenamefont {Katsumi}, \citenamefont {Massat}, \citenamefont
  {Wen}, \citenamefont {Chen}, \citenamefont {Gallais},\ and\ \citenamefont
  {Shimano}}]{Grasset2022NematicMode}%
  \BibitemOpen
  \bibfield  {author} {\bibinfo {author} {\bibfnamefont {R.}~\bibnamefont
  {Grasset}}, \bibinfo {author} {\bibfnamefont {K.}~\bibnamefont {Katsumi}},
  \bibinfo {author} {\bibfnamefont {P.}~\bibnamefont {Massat}}, \bibinfo
  {author} {\bibfnamefont {H.-H.}\ \bibnamefont {Wen}}, \bibinfo {author}
  {\bibfnamefont {X.-H.}\ \bibnamefont {Chen}}, \bibinfo {author}
  {\bibfnamefont {Y.}~\bibnamefont {Gallais}},\ and\ \bibinfo {author}
  {\bibfnamefont {R.}~\bibnamefont {Shimano}},\ }\href
  {https://doi.org/10.1038/s41535-021-00411-9} {\bibfield  {journal} {\bibinfo
  {journal} {npj Quantum Materials}\ }\textbf {\bibinfo {volume} {7}},\
  \bibinfo {pages} {4} (\bibinfo {year} {2022})}\BibitemShut {NoStop}%
\bibitem [{\citenamefont {Perez-Salinas}\ \emph {et~al.}(2022)\citenamefont
  {Perez-Salinas}, \citenamefont {Johnson}, \citenamefont {Prabhakaran},\ and\
  \citenamefont {Wall}}]{PerezSalinas2022Multimode}%
  \BibitemOpen
  \bibfield  {author} {\bibinfo {author} {\bibfnamefont {D.}~\bibnamefont
  {Perez-Salinas}}, \bibinfo {author} {\bibfnamefont {A.~S.}\ \bibnamefont
  {Johnson}}, \bibinfo {author} {\bibfnamefont {D.}~\bibnamefont
  {Prabhakaran}},\ and\ \bibinfo {author} {\bibfnamefont {S.}~\bibnamefont
  {Wall}},\ }\href {https://doi.org/10.1038/s41467-021-27819-y} {\bibfield
  {journal} {\bibinfo  {journal} {Nature Communications}\ }\textbf {\bibinfo
  {volume} {13}},\ \bibinfo {pages} {238} (\bibinfo {year} {2022})}\BibitemShut
  {NoStop}%
\bibitem [{\citenamefont {de~la Torre}\ \emph {et~al.}(2022)\citenamefont
  {de~la Torre}, \citenamefont {Seyler}, \citenamefont {Buchhold},
  \citenamefont {Baum}, \citenamefont {Zhang}, \citenamefont {Laurita},
  \citenamefont {Harter}, \citenamefont {Zhao}, \citenamefont {Phinney},
  \citenamefont {Chen}, \citenamefont {Wilson}, \citenamefont {Cao},
  \citenamefont {Averitt}, \citenamefont {Refael},\ and\ \citenamefont
  {Hsieh}}]{delaTorre2022DrivenMott}%
  \BibitemOpen
  \bibfield  {author} {\bibinfo {author} {\bibfnamefont {A.}~\bibnamefont
  {de~la Torre}}, \bibinfo {author} {\bibfnamefont {K.~L.}\ \bibnamefont
  {Seyler}}, \bibinfo {author} {\bibfnamefont {M.}~\bibnamefont {Buchhold}},
  \bibinfo {author} {\bibfnamefont {Y.}~\bibnamefont {Baum}}, \bibinfo {author}
  {\bibfnamefont {G.}~\bibnamefont {Zhang}}, \bibinfo {author} {\bibfnamefont
  {N.~J.}\ \bibnamefont {Laurita}}, \bibinfo {author} {\bibfnamefont {J.~W.}\
  \bibnamefont {Harter}}, \bibinfo {author} {\bibfnamefont {L.}~\bibnamefont
  {Zhao}}, \bibinfo {author} {\bibfnamefont {I.}~\bibnamefont {Phinney}},
  \bibinfo {author} {\bibfnamefont {X.}~\bibnamefont {Chen}}, \bibinfo {author}
  {\bibfnamefont {S.~D.}\ \bibnamefont {Wilson}}, \bibinfo {author}
  {\bibfnamefont {G.}~\bibnamefont {Cao}}, \bibinfo {author} {\bibfnamefont
  {R.~D.}\ \bibnamefont {Averitt}}, \bibinfo {author} {\bibfnamefont
  {G.}~\bibnamefont {Refael}},\ and\ \bibinfo {author} {\bibfnamefont
  {D.}~\bibnamefont {Hsieh}},\ }\href
  {https://doi.org/10.1038/s42005-022-00813-6} {\bibfield  {journal} {\bibinfo
  {journal} {Communications Physics}\ }\textbf {\bibinfo {volume} {5}},\
  \bibinfo {pages} {35} (\bibinfo {year} {2022})}\BibitemShut {NoStop}%
\bibitem [{\citenamefont {Fausti}\ \emph {et~al.}(2011)\citenamefont {Fausti},
  \citenamefont {Tobey}, \citenamefont {Dean}, \citenamefont {Kaiser},
  \citenamefont {Dienst}, \citenamefont {Hoffmann}, \citenamefont {Pyon},
  \citenamefont {Takayama}, \citenamefont {Takagi},\ and\ \citenamefont
  {Cavalleri}}]{Fausti2011LightInduced}%
  \BibitemOpen
  \bibfield  {author} {\bibinfo {author} {\bibfnamefont {D.}~\bibnamefont
  {Fausti}}, \bibinfo {author} {\bibfnamefont {R.~I.}\ \bibnamefont {Tobey}},
  \bibinfo {author} {\bibfnamefont {N.}~\bibnamefont {Dean}}, \bibinfo {author}
  {\bibfnamefont {S.}~\bibnamefont {Kaiser}}, \bibinfo {author} {\bibfnamefont
  {A.}~\bibnamefont {Dienst}}, \bibinfo {author} {\bibfnamefont {M.~C.}\
  \bibnamefont {Hoffmann}}, \bibinfo {author} {\bibfnamefont {S.}~\bibnamefont
  {Pyon}}, \bibinfo {author} {\bibfnamefont {T.}~\bibnamefont {Takayama}},
  \bibinfo {author} {\bibfnamefont {H.}~\bibnamefont {Takagi}},\ and\ \bibinfo
  {author} {\bibfnamefont {A.}~\bibnamefont {Cavalleri}},\ }\href
  {https://doi.org/10.1126/science.1197294} {\bibfield  {journal} {\bibinfo
  {journal} {Science}\ }\textbf {\bibinfo {volume} {331}},\ \bibinfo {pages}
  {189} (\bibinfo {year} {2011})}\BibitemShut {NoStop}%
\bibitem [{\citenamefont {Hu}\ \emph {et~al.}(2014)\citenamefont {Hu},
  \citenamefont {Kaiser}, \citenamefont {Nicoletti}, \citenamefont {Hunt},
  \citenamefont {Gierz}, \citenamefont {Hoffmann}, \citenamefont {Le~Tacon},
  \citenamefont {Loew}, \citenamefont {Keimer},\ and\ \citenamefont
  {Cavalleri}}]{Hu2014OpticallyEnhanced}%
  \BibitemOpen
  \bibfield  {author} {\bibinfo {author} {\bibfnamefont {W.}~\bibnamefont
  {Hu}}, \bibinfo {author} {\bibfnamefont {S.}~\bibnamefont {Kaiser}}, \bibinfo
  {author} {\bibfnamefont {D.}~\bibnamefont {Nicoletti}}, \bibinfo {author}
  {\bibfnamefont {C.~R.}\ \bibnamefont {Hunt}}, \bibinfo {author}
  {\bibfnamefont {I.}~\bibnamefont {Gierz}}, \bibinfo {author} {\bibfnamefont
  {M.~C.}\ \bibnamefont {Hoffmann}}, \bibinfo {author} {\bibfnamefont
  {M.}~\bibnamefont {Le~Tacon}}, \bibinfo {author} {\bibfnamefont
  {T.}~\bibnamefont {Loew}}, \bibinfo {author} {\bibfnamefont {B.}~\bibnamefont
  {Keimer}},\ and\ \bibinfo {author} {\bibfnamefont {A.}~\bibnamefont
  {Cavalleri}},\ }\href {https://doi.org/10.1038/nmat3963} {\bibfield
  {journal} {\bibinfo  {journal} {Nature Materials}\ }\textbf {\bibinfo
  {volume} {13}},\ \bibinfo {pages} {705} (\bibinfo {year} {2014})}\BibitemShut
  {NoStop}%
\bibitem [{\citenamefont {Kaiser}\ \emph {et~al.}(2014)\citenamefont {Kaiser},
  \citenamefont {Hunt}, \citenamefont {Nicoletti}, \citenamefont {Hu},
  \citenamefont {Gierz}, \citenamefont {Liu}, \citenamefont {Le~Tacon},
  \citenamefont {Loew}, \citenamefont {Haug}, \citenamefont {Keimer},\ and\
  \citenamefont {Cavalleri}}]{Kaiser2014Optically}%
  \BibitemOpen
  \bibfield  {author} {\bibinfo {author} {\bibfnamefont {S.}~\bibnamefont
  {Kaiser}}, \bibinfo {author} {\bibfnamefont {C.~R.}\ \bibnamefont {Hunt}},
  \bibinfo {author} {\bibfnamefont {D.}~\bibnamefont {Nicoletti}}, \bibinfo
  {author} {\bibfnamefont {W.}~\bibnamefont {Hu}}, \bibinfo {author}
  {\bibfnamefont {I.}~\bibnamefont {Gierz}}, \bibinfo {author} {\bibfnamefont
  {H.~Y.}\ \bibnamefont {Liu}}, \bibinfo {author} {\bibfnamefont
  {M.}~\bibnamefont {Le~Tacon}}, \bibinfo {author} {\bibfnamefont
  {T.}~\bibnamefont {Loew}}, \bibinfo {author} {\bibfnamefont {D.}~\bibnamefont
  {Haug}}, \bibinfo {author} {\bibfnamefont {B.}~\bibnamefont {Keimer}},\ and\
  \bibinfo {author} {\bibfnamefont {A.}~\bibnamefont {Cavalleri}},\ }\href
  {https://doi.org/10.1103/PhysRevB.89.184516} {\bibfield  {journal} {\bibinfo
  {journal} {Physical Review B}\ }\textbf {\bibinfo {volume} {89}},\ \bibinfo
  {pages} {184516} (\bibinfo {year} {2014})}\BibitemShut {NoStop}%
\bibitem [{\citenamefont {Mitrano}\ \emph {et~al.}(2016)\citenamefont
  {Mitrano}, \citenamefont {Cantaluppi}, \citenamefont {Nicoletti},
  \citenamefont {Kaiser}, \citenamefont {Perucchi}, \citenamefont {Lupi},
  \citenamefont {Di~Pietro}, \citenamefont {Pontiroli}, \citenamefont
  {Ricc\`o}, \citenamefont {Clark}, \citenamefont {Jaksch},\ and\ \citenamefont
  {Cavalleri}}]{Mittrano2016Possible}%
  \BibitemOpen
  \bibfield  {author} {\bibinfo {author} {\bibfnamefont {M.}~\bibnamefont
  {Mitrano}}, \bibinfo {author} {\bibfnamefont {A.}~\bibnamefont {Cantaluppi}},
  \bibinfo {author} {\bibfnamefont {D.}~\bibnamefont {Nicoletti}}, \bibinfo
  {author} {\bibfnamefont {S.}~\bibnamefont {Kaiser}}, \bibinfo {author}
  {\bibfnamefont {A.}~\bibnamefont {Perucchi}}, \bibinfo {author}
  {\bibfnamefont {S.}~\bibnamefont {Lupi}}, \bibinfo {author} {\bibfnamefont
  {P.}~\bibnamefont {Di~Pietro}}, \bibinfo {author} {\bibfnamefont
  {D.}~\bibnamefont {Pontiroli}}, \bibinfo {author} {\bibfnamefont
  {M.}~\bibnamefont {Ricc\`o}}, \bibinfo {author} {\bibfnamefont {S.~R.}\
  \bibnamefont {Clark}}, \bibinfo {author} {\bibfnamefont {D.}~\bibnamefont
  {Jaksch}},\ and\ \bibinfo {author} {\bibfnamefont {A.}~\bibnamefont
  {Cavalleri}},\ }\href {https://doi.org/10.1038/nature16522} {\bibfield
  {journal} {\bibinfo  {journal} {Nature}\ }\textbf {\bibinfo {volume} {530}},\
  \bibinfo {pages} {461} (\bibinfo {year} {2016})}\BibitemShut {NoStop}%
\bibitem [{\citenamefont {Cavalleri}(2018)}]{Cavalleri2018Photo}%
  \BibitemOpen
  \bibfield  {author} {\bibinfo {author} {\bibfnamefont {A.}~\bibnamefont
  {Cavalleri}},\ }\href {https://doi.org/10.1080/00107514.2017.1406623}
  {\bibfield  {journal} {\bibinfo  {journal} {Contemporary Physics}\ }\textbf
  {\bibinfo {volume} {59}},\ \bibinfo {pages} {31} (\bibinfo {year}
  {2018})}\BibitemShut {NoStop}%
\bibitem [{\citenamefont {Budden}\ \emph {et~al.}(2021)\citenamefont {Budden},
  \citenamefont {Gebert}, \citenamefont {Buzzi}, \citenamefont {Jotzu},
  \citenamefont {Wang}, \citenamefont {Matsuyama}, \citenamefont {Meier},
  \citenamefont {Laplace}, \citenamefont {Pontiroli}, \citenamefont {Ricc\`o}
  \emph {et~al.}}]{Budden2021Evidence}%
  \BibitemOpen
  \bibfield  {author} {\bibinfo {author} {\bibfnamefont {M.}~\bibnamefont
  {Budden}}, \bibinfo {author} {\bibfnamefont {T.}~\bibnamefont {Gebert}},
  \bibinfo {author} {\bibfnamefont {M.}~\bibnamefont {Buzzi}}, \bibinfo
  {author} {\bibfnamefont {G.}~\bibnamefont {Jotzu}}, \bibinfo {author}
  {\bibfnamefont {E.}~\bibnamefont {Wang}}, \bibinfo {author} {\bibfnamefont
  {T.}~\bibnamefont {Matsuyama}}, \bibinfo {author} {\bibfnamefont
  {G.}~\bibnamefont {Meier}}, \bibinfo {author} {\bibfnamefont
  {Y.}~\bibnamefont {Laplace}}, \bibinfo {author} {\bibfnamefont
  {D.}~\bibnamefont {Pontiroli}}, \bibinfo {author} {\bibfnamefont
  {M.}~\bibnamefont {Ricc\`o}}, \emph {et~al.},\ }\href
  {https://doi.org/10.1038/s41567-020-01148-1} {\bibfield  {journal} {\bibinfo
  {journal} {Nature Physics}\ }\textbf {\bibinfo {volume} {17}},\ \bibinfo
  {pages} {611} (\bibinfo {year} {2021})}\BibitemShut {NoStop}%
\bibitem [{\citenamefont {Buzzi}\ \emph {et~al.}(2020)\citenamefont {Buzzi},
  \citenamefont {Nicoletti}, \citenamefont {Fechner}, \citenamefont
  {Tancogne-Dejean}, \citenamefont {Sentef}, \citenamefont {Georges},\ and\
  \citenamefont {Cavalleri}}]{Buzzi2020Photomolecular}%
  \BibitemOpen
  \bibfield  {author} {\bibinfo {author} {\bibfnamefont {M.}~\bibnamefont
  {Buzzi}}, \bibinfo {author} {\bibfnamefont {D.}~\bibnamefont {Nicoletti}},
  \bibinfo {author} {\bibfnamefont {M.}~\bibnamefont {Fechner}}, \bibinfo
  {author} {\bibfnamefont {N.}~\bibnamefont {Tancogne-Dejean}}, \bibinfo
  {author} {\bibfnamefont {M.~A.}\ \bibnamefont {Sentef}}, \bibinfo {author}
  {\bibfnamefont {A.}~\bibnamefont {Georges}},\ and\ \bibinfo {author}
  {\bibfnamefont {A.}~\bibnamefont {Cavalleri}},\ }\href
  {https://doi.org/10.1103/PhysRevX.10.031028} {\bibfield  {journal} {\bibinfo
  {journal} {Physical Review X}\ }\textbf {\bibinfo {volume} {10}},\ \bibinfo
  {pages} {031028} (\bibinfo {year} {2020})}\BibitemShut {NoStop}%
\bibitem [{\citenamefont {Liu}\ \emph {et~al.}(2020)\citenamefont {Liu},
  \citenamefont {F\"orst}, \citenamefont {Fechner}, \citenamefont {Nicoletti},
  \citenamefont {Porras}, \citenamefont {Loew}, \citenamefont {Keimer},\ and\
  \citenamefont {Cavalleri}}]{Liu2020PumpFrequencyResonances}%
  \BibitemOpen
  \bibfield  {author} {\bibinfo {author} {\bibfnamefont {B.}~\bibnamefont
  {Liu}}, \bibinfo {author} {\bibfnamefont {M.}~\bibnamefont {F\"orst}},
  \bibinfo {author} {\bibfnamefont {M.}~\bibnamefont {Fechner}}, \bibinfo
  {author} {\bibfnamefont {D.}~\bibnamefont {Nicoletti}}, \bibinfo {author}
  {\bibfnamefont {J.}~\bibnamefont {Porras}}, \bibinfo {author} {\bibfnamefont
  {T.}~\bibnamefont {Loew}}, \bibinfo {author} {\bibfnamefont {B.}~\bibnamefont
  {Keimer}},\ and\ \bibinfo {author} {\bibfnamefont {A.}~\bibnamefont
  {Cavalleri}},\ }\href {https://doi.org/10.1103/PhysRevX.10.011053} {\bibfield
   {journal} {\bibinfo  {journal} {Phys. Rev. X}\ }\textbf {\bibinfo {volume}
  {10}},\ \bibinfo {pages} {011053} (\bibinfo {year} {2020})}\BibitemShut
  {NoStop}%
\bibitem [{\citenamefont {Fava}\ \emph {et~al.}(2024)\citenamefont {Fava},
  \citenamefont {De~Vecchi}, \citenamefont {Jotzu}, \citenamefont {Buzzi},
  \citenamefont {Gebert}, \citenamefont {Liu}, \citenamefont {Keimer},\ and\
  \citenamefont {Cavalleri}}]{Fava2024Magnetic}%
  \BibitemOpen
  \bibfield  {author} {\bibinfo {author} {\bibfnamefont {S.}~\bibnamefont
  {Fava}}, \bibinfo {author} {\bibfnamefont {G.}~\bibnamefont {De~Vecchi}},
  \bibinfo {author} {\bibfnamefont {G.}~\bibnamefont {Jotzu}}, \bibinfo
  {author} {\bibfnamefont {M.}~\bibnamefont {Buzzi}}, \bibinfo {author}
  {\bibfnamefont {T.}~\bibnamefont {Gebert}}, \bibinfo {author} {\bibfnamefont
  {Y.}~\bibnamefont {Liu}}, \bibinfo {author} {\bibfnamefont {B.}~\bibnamefont
  {Keimer}},\ and\ \bibinfo {author} {\bibfnamefont {A.}~\bibnamefont
  {Cavalleri}},\ }\href {https://doi.org/10.1038/s41586-024-07635-2} {\bibfield
   {journal} {\bibinfo  {journal} {Nature}\ }\textbf {\bibinfo {volume}
  {632}},\ \bibinfo {pages} {75} (\bibinfo {year} {2024})}\BibitemShut
  {NoStop}%
\bibitem [{\citenamefont {Rosenberg}\ \emph {et~al.}(2025)\citenamefont
  {Rosenberg}, \citenamefont {Nicoletti}, \citenamefont {Buzzi}, \citenamefont
  {Iudica}, \citenamefont {Putzke}, \citenamefont {Liu}, \citenamefont
  {Keimer},\ and\ \citenamefont {Cavalleri}}]{Rosenberg2025Signatures}%
  \BibitemOpen
  \bibfield  {author} {\bibinfo {author} {\bibfnamefont {M.}~\bibnamefont
  {Rosenberg}}, \bibinfo {author} {\bibfnamefont {D.}~\bibnamefont
  {Nicoletti}}, \bibinfo {author} {\bibfnamefont {M.}~\bibnamefont {Buzzi}},
  \bibinfo {author} {\bibfnamefont {A.}~\bibnamefont {Iudica}}, \bibinfo
  {author} {\bibfnamefont {C.}~\bibnamefont {Putzke}}, \bibinfo {author}
  {\bibfnamefont {Y.}~\bibnamefont {Liu}}, \bibinfo {author} {\bibfnamefont
  {B.}~\bibnamefont {Keimer}},\ and\ \bibinfo {author} {\bibfnamefont
  {A.}~\bibnamefont {Cavalleri}},\ }\href {https://doi.org/10.1103/2m3d-s3j9}
  {\bibfield  {journal} {\bibinfo  {journal} {Physical Review B}\ }\textbf
  {\bibinfo {volume} {112}},\ \bibinfo {pages} {214522} (\bibinfo {year}
  {2025})}\BibitemShut {NoStop}%
\bibitem [{\citenamefont {Michael}\ \emph
  {et~al.}(2026{\natexlab{a}})\citenamefont {Michael}, \citenamefont
  {De~Santis}, \citenamefont {Demler},\ and\ \citenamefont
  {Lee}}]{Michael2026FluxFloquet}%
  \BibitemOpen
  \bibfield  {author} {\bibinfo {author} {\bibfnamefont {M.~H.}\ \bibnamefont
  {Michael}}, \bibinfo {author} {\bibfnamefont {D.}~\bibnamefont {De~Santis}},
  \bibinfo {author} {\bibfnamefont {E.~A.}\ \bibnamefont {Demler}},\ and\
  \bibinfo {author} {\bibfnamefont {P.~A.}\ \bibnamefont {Lee}},\ }\href
  {https://doi.org/10.1103/tq9r-g727} {\bibfield  {journal} {\bibinfo
  {journal} {Phys. Rev. X}\ }\textbf {\bibinfo {volume} {16}},\ \bibinfo
  {pages} {031055} (\bibinfo {year} {2026}{\natexlab{a}})}\BibitemShut
  {NoStop}%
\bibitem [{\citenamefont {Zhang}\ and\ \citenamefont
  {Fertig}(2005)}]{Zhang2005Vortices}%
  \BibitemOpen
  \bibfield  {author} {\bibinfo {author} {\bibfnamefont {W.}~\bibnamefont
  {Zhang}}\ and\ \bibinfo {author} {\bibfnamefont {H.~A.}\ \bibnamefont
  {Fertig}},\ }\href {https://doi.org/10.1103/PhysRevB.71.224514} {\bibfield
  {journal} {\bibinfo  {journal} {Phys. Rev. B}\ }\textbf {\bibinfo {volume}
  {71}},\ \bibinfo {pages} {224514} (\bibinfo {year} {2005})}\BibitemShut
  {NoStop}%
\bibitem [{\citenamefont {Homann}\ \emph {et~al.}(2024)\citenamefont {Homann},
  \citenamefont {Michael}, \citenamefont {Cosme},\ and\ \citenamefont
  {Mathey}}]{Homann2024DissipationlessCounterflow}%
  \BibitemOpen
  \bibfield  {author} {\bibinfo {author} {\bibfnamefont {G.}~\bibnamefont
  {Homann}}, \bibinfo {author} {\bibfnamefont {M.~H.}\ \bibnamefont {Michael}},
  \bibinfo {author} {\bibfnamefont {J.~G.}\ \bibnamefont {Cosme}},\ and\
  \bibinfo {author} {\bibfnamefont {L.}~\bibnamefont {Mathey}},\ }\href
  {https://doi.org/10.1103/PhysRevLett.132.096002} {\bibfield  {journal}
  {\bibinfo  {journal} {Phys. Rev. Lett.}\ }\textbf {\bibinfo {volume} {132}},\
  \bibinfo {pages} {096002} (\bibinfo {year} {2024})}\BibitemShut {NoStop}%
\bibitem [{\citenamefont {Michael}\ \emph
  {et~al.}(2026{\natexlab{b}})\citenamefont {Michael}, \citenamefont
  {De~Santis}, \citenamefont {Demler},\ and\ \citenamefont
  {Lee}}]{Michael2026Counterflow}%
  \BibitemOpen
  \bibfield  {author} {\bibinfo {author} {\bibfnamefont {M.~H.}\ \bibnamefont
  {Michael}}, \bibinfo {author} {\bibfnamefont {D.}~\bibnamefont {De~Santis}},
  \bibinfo {author} {\bibfnamefont {E.~A.}\ \bibnamefont {Demler}},\ and\
  \bibinfo {author} {\bibfnamefont {P.~A.}\ \bibnamefont {Lee}},\ }\href@noop
  {} {\bibinfo {title} {Manuscript in preparation}} (\bibinfo {year}
  {2026}{\natexlab{b}})\BibitemShut {NoStop}%
\bibitem [{\citenamefont {{De Santis}}\ \emph {et~al.}(2025)\citenamefont {{De
  Santis}}, \citenamefont {Michael}, \citenamefont {Chattopadhyay},
  \citenamefont {Cavalleri}, \citenamefont {Refael}, \citenamefont {Lee},\ and\
  \citenamefont {Demler}}]{DeSantis2025Enhanced}%
  \BibitemOpen
  \bibfield  {author} {\bibinfo {author} {\bibfnamefont {D.}~\bibnamefont {{De
  Santis}}}, \bibinfo {author} {\bibfnamefont {M.~H.}\ \bibnamefont {Michael}},
  \bibinfo {author} {\bibfnamefont {S.}~\bibnamefont {Chattopadhyay}}, \bibinfo
  {author} {\bibfnamefont {A.}~\bibnamefont {Cavalleri}}, \bibinfo {author}
  {\bibfnamefont {G.}~\bibnamefont {Refael}}, \bibinfo {author} {\bibfnamefont
  {P.~A.}\ \bibnamefont {Lee}},\ and\ \bibinfo {author} {\bibfnamefont {E.~A.}\
  \bibnamefont {Demler}},\ }\href {https://doi.org/10.48550/arXiv.2511.12287}
  {\bibinfo {title} {Enhanced coherence in the periodically driven
  two-dimensional {XY} model}} (\bibinfo {year} {2025}),\ \Eprint
  {https://arxiv.org/abs/2511.12287} {arXiv:2511.12287 [cond-mat.supr-con]}
  \BibitemShut {NoStop}%
\bibitem [{\citenamefont {Diessel}\ \emph {et~al.}(2026)\citenamefont
  {Diessel}, \citenamefont {Sachdev},\ and\ \citenamefont
  {Bonetti}}]{Diessel2026SteadyStates}%
  \BibitemOpen
  \bibfield  {author} {\bibinfo {author} {\bibfnamefont {O.~K.}\ \bibnamefont
  {Diessel}}, \bibinfo {author} {\bibfnamefont {S.}~\bibnamefont {Sachdev}},\
  and\ \bibinfo {author} {\bibfnamefont {P.~M.}\ \bibnamefont {Bonetti}},\
  }\href {https://doi.org/10.1088/1361-6633/ae4ad3} {\bibfield  {journal}
  {\bibinfo  {journal} {Reports on Progress in Physics}\ }\textbf {\bibinfo
  {volume} {89}},\ \bibinfo {pages} {038001} (\bibinfo {year}
  {2026})}\BibitemShut {NoStop}%
\bibitem [{\citenamefont {Mankowsky}\ \emph {et~al.}(2014)\citenamefont
  {Mankowsky}, \citenamefont {Subedi}, \citenamefont {F\"orst}, \citenamefont
  {Mariager}, \citenamefont {Chollet}, \citenamefont {Lemke}, \citenamefont
  {Robinson}, \citenamefont {Glownia}, \citenamefont {Minitti}, \citenamefont
  {Frano} \emph {et~al.}}]{Mankowsky2014Nonlinear}%
  \BibitemOpen
  \bibfield  {author} {\bibinfo {author} {\bibfnamefont {R.}~\bibnamefont
  {Mankowsky}}, \bibinfo {author} {\bibfnamefont {A.}~\bibnamefont {Subedi}},
  \bibinfo {author} {\bibfnamefont {M.}~\bibnamefont {F\"orst}}, \bibinfo
  {author} {\bibfnamefont {S.~O.}\ \bibnamefont {Mariager}}, \bibinfo {author}
  {\bibfnamefont {M.}~\bibnamefont {Chollet}}, \bibinfo {author} {\bibfnamefont
  {H.~T.}\ \bibnamefont {Lemke}}, \bibinfo {author} {\bibfnamefont {J.~S.}\
  \bibnamefont {Robinson}}, \bibinfo {author} {\bibfnamefont {J.~M.}\
  \bibnamefont {Glownia}}, \bibinfo {author} {\bibfnamefont {M.~P.}\
  \bibnamefont {Minitti}}, \bibinfo {author} {\bibfnamefont {A.}~\bibnamefont
  {Frano}}, \emph {et~al.},\ }\href {https://doi.org/10.1038/nature13875}
  {\bibfield  {journal} {\bibinfo  {journal} {Nature}\ }\textbf {\bibinfo
  {volume} {516}},\ \bibinfo {pages} {71} (\bibinfo {year} {2014})}\BibitemShut
  {NoStop}%
\bibitem [{\citenamefont {Fechner}\ and\ \citenamefont
  {Spaldin}(2016)}]{Fechner2016EffectsOfIntense}%
  \BibitemOpen
  \bibfield  {author} {\bibinfo {author} {\bibfnamefont {M.}~\bibnamefont
  {Fechner}}\ and\ \bibinfo {author} {\bibfnamefont {N.~A.}\ \bibnamefont
  {Spaldin}},\ }\href {https://doi.org/10.1103/PhysRevB.94.134307} {\bibfield
  {journal} {\bibinfo  {journal} {Phys. Rev. B}\ }\textbf {\bibinfo {volume}
  {94}},\ \bibinfo {pages} {134307} (\bibinfo {year} {2016})}\BibitemShut
  {NoStop}%
\bibitem [{\citenamefont {Dubroka}\ \emph {et~al.}(2011)\citenamefont
  {Dubroka}, \citenamefont {R\"ossle}, \citenamefont {Kim}, \citenamefont
  {Malik}, \citenamefont {Munzar}, \citenamefont {Basov}, \citenamefont
  {Schafgans}, \citenamefont {Moon}, \citenamefont {Lin}, \citenamefont {Haug},
  \citenamefont {Hinkov}, \citenamefont {Keimer}, \citenamefont {Wolf},
  \citenamefont {Storey}, \citenamefont {Tallon},\ and\ \citenamefont
  {Bernhard}}]{Dubroka2011Evidence}%
  \BibitemOpen
  \bibfield  {author} {\bibinfo {author} {\bibfnamefont {A.}~\bibnamefont
  {Dubroka}}, \bibinfo {author} {\bibfnamefont {M.}~\bibnamefont {R\"ossle}},
  \bibinfo {author} {\bibfnamefont {K.~W.}\ \bibnamefont {Kim}}, \bibinfo
  {author} {\bibfnamefont {V.~K.}\ \bibnamefont {Malik}}, \bibinfo {author}
  {\bibfnamefont {D.}~\bibnamefont {Munzar}}, \bibinfo {author} {\bibfnamefont
  {D.~N.}\ \bibnamefont {Basov}}, \bibinfo {author} {\bibfnamefont {A.~A.}\
  \bibnamefont {Schafgans}}, \bibinfo {author} {\bibfnamefont {S.~J.}\
  \bibnamefont {Moon}}, \bibinfo {author} {\bibfnamefont {C.~T.}\ \bibnamefont
  {Lin}}, \bibinfo {author} {\bibfnamefont {D.}~\bibnamefont {Haug}}, \bibinfo
  {author} {\bibfnamefont {V.}~\bibnamefont {Hinkov}}, \bibinfo {author}
  {\bibfnamefont {B.}~\bibnamefont {Keimer}}, \bibinfo {author} {\bibfnamefont
  {T.}~\bibnamefont {Wolf}}, \bibinfo {author} {\bibfnamefont {J.~G.}\
  \bibnamefont {Storey}}, \bibinfo {author} {\bibfnamefont {J.~L.}\
  \bibnamefont {Tallon}},\ and\ \bibinfo {author} {\bibfnamefont
  {C.}~\bibnamefont {Bernhard}},\ }\href
  {https://doi.org/10.1103/PhysRevLett.106.047006} {\bibfield  {journal}
  {\bibinfo  {journal} {Phys. Rev. Lett.}\ }\textbf {\bibinfo {volume} {106}},\
  \bibinfo {pages} {047006} (\bibinfo {year} {2011})}\BibitemShut {NoStop}%
\bibitem [{\citenamefont {Uykur}\ \emph {et~al.}(2014)\citenamefont {Uykur},
  \citenamefont {Tanaka}, \citenamefont {Masui}, \citenamefont {Miyasaka},\
  and\ \citenamefont {Tajima}}]{Uykur2014Persistence}%
  \BibitemOpen
  \bibfield  {author} {\bibinfo {author} {\bibfnamefont {E.}~\bibnamefont
  {Uykur}}, \bibinfo {author} {\bibfnamefont {K.}~\bibnamefont {Tanaka}},
  \bibinfo {author} {\bibfnamefont {T.}~\bibnamefont {Masui}}, \bibinfo
  {author} {\bibfnamefont {S.}~\bibnamefont {Miyasaka}},\ and\ \bibinfo
  {author} {\bibfnamefont {S.}~\bibnamefont {Tajima}},\ }\href
  {https://doi.org/10.1103/PhysRevLett.112.127003} {\bibfield  {journal}
  {\bibinfo  {journal} {Phys. Rev. Lett.}\ }\textbf {\bibinfo {volume} {112}},\
  \bibinfo {pages} {127003} (\bibinfo {year} {2014})}\BibitemShut {NoStop}%
\bibitem [{\citenamefont {Lawrence}\ and\ \citenamefont
  {Doniach}(1971)}]{Lawrence1971LayerStructureSuperconductors}%
  \BibitemOpen
  \bibfield  {author} {\bibinfo {author} {\bibfnamefont {W.~E.}\ \bibnamefont
  {Lawrence}}\ and\ \bibinfo {author} {\bibfnamefont {S.}~\bibnamefont
  {Doniach}},\ }in\ \href@noop {} {\emph {\bibinfo {booktitle} {Proceedings of
  the Twelfth International Conference on Low Temperature Physics}}},\ \bibinfo
  {editor} {edited by\ \bibinfo {editor} {\bibfnamefont {E.}~\bibnamefont
  {Kanda}}}\ (\bibinfo  {publisher} {Keigaku Publishing Co., Ltd.},\ \bibinfo
  {address} {Tokyo},\ \bibinfo {year} {1971})\ pp.\ \bibinfo {pages}
  {361--362}\BibitemShut {NoStop}%
\bibitem [{\citenamefont {Bulaevskii}\ \emph {et~al.}(1992)\citenamefont
  {Bulaevskii}, \citenamefont {Ledvij},\ and\ \citenamefont
  {Kogan}}]{Bulaevskii1992Vortices}%
  \BibitemOpen
  \bibfield  {author} {\bibinfo {author} {\bibfnamefont {L.~N.}\ \bibnamefont
  {Bulaevskii}}, \bibinfo {author} {\bibfnamefont {M.}~\bibnamefont {Ledvij}},\
  and\ \bibinfo {author} {\bibfnamefont {V.~G.}\ \bibnamefont {Kogan}},\ }\href
  {https://doi.org/10.1103/PhysRevB.46.366} {\bibfield  {journal} {\bibinfo
  {journal} {Phys. Rev. B}\ }\textbf {\bibinfo {volume} {46}},\ \bibinfo
  {pages} {366} (\bibinfo {year} {1992})}\BibitemShut {NoStop}%
\bibitem [{Note1()}]{Note1}%
  \BibitemOpen
  \bibinfo {note} {A superscript $*$ denotes a dimensionless quantity. We use
  $H^{*}=H/J_0$, $J_\ell ^{*}=J_\ell /J_0$, $J_\perp ^{*}=J_\perp /J_0$,
  $T^{*}=k_{\protect \rm B}T/J_0$, $t^{*}=t/\tau =\Gamma J_0t$, and $\Omega
  ^{*}=\Omega \tau $, with $\tau =(\Gamma J_0)^{-1}$. Here $\Gamma $ is the
  kinetic coefficient that enters in the model-A dynamical equation. The
  Peierls phase $A_{\ell ,ij}^{*}=(2e/\hbar )\DOTSI \intop \ilimits@
  _i^j\protect \mathbf A_\ell \protect \!\cdot \protect \mathrm {d}\protect \bm
  {\ell }$ is dimensionless as well. To lighten the notation, we drop the
  superscript $*$ throughout the remainder of the main text.}\BibitemShut
  {Stop}%
\bibitem [{Note2()}]{Note2}%
  \BibitemOpen
  \bibinfo {note} {For an anisotropic phase-only model, the ratio of interlayer
  to in-plane couplings can be estimated from the zero-temperature penetration
  depths. We write~\cite {Mihlin2009TemperatureDependence} $ J_\perp \sim \left
  ( \protect \frac {\lambda _{ab} \protect \tmspace +\thickmuskip {.2777em}
  a}{\lambda _c \protect \tmspace +\thickmuskip {.2777em} d_\perp } \right )^2,
  $ where $a\simeq 3.8\protect \,$\r A{} is the in-plane lattice constant and
  $d_\perp \simeq 4\protect \,$\r A{} is the intrabilayer separation. Using the
  YBCO penetration-depth values quoted in Ref.~\cite
  {Mihlin2009TemperatureDependence} gives an intrabilayer coupling ratio of
  order $ J_\perp \sim 10^{-3}\protect \text {--}10^{-2}. $ We therefore use
  $J_\perp \sim 0.01$ as a representative weak-coupling value near the upper
  end of this experimentally motivated range.}\BibitemShut {Stop}%
\bibitem [{\citenamefont {Hohenberg}\ and\ \citenamefont
  {Halperin}(1977)}]{Hohenberg1977Theory}%
  \BibitemOpen
  \bibfield  {author} {\bibinfo {author} {\bibfnamefont {P.~C.}\ \bibnamefont
  {Hohenberg}}\ and\ \bibinfo {author} {\bibfnamefont {B.~I.}\ \bibnamefont
  {Halperin}},\ }\href {https://doi.org/10.1103/RevModPhys.49.435} {\bibfield
  {journal} {\bibinfo  {journal} {Reviews of Modern Physics}\ }\textbf
  {\bibinfo {volume} {49}},\ \bibinfo {pages} {435} (\bibinfo {year}
  {1977})}\BibitemShut {NoStop}%
\bibitem [{\citenamefont {Podolsky}\ \emph {et~al.}(2007)\citenamefont
  {Podolsky}, \citenamefont {Raghu},\ and\ \citenamefont
  {Vishwanath}}]{Podolsky2007Nernst}%
  \BibitemOpen
  \bibfield  {author} {\bibinfo {author} {\bibfnamefont {D.}~\bibnamefont
  {Podolsky}}, \bibinfo {author} {\bibfnamefont {S.}~\bibnamefont {Raghu}},\
  and\ \bibinfo {author} {\bibfnamefont {A.}~\bibnamefont {Vishwanath}},\
  }\href {https://doi.org/10.1103/PhysRevLett.99.117004} {\bibfield  {journal}
  {\bibinfo  {journal} {Physical Review Letters}\ }\textbf {\bibinfo {volume}
  {99}},\ \bibinfo {pages} {117004} (\bibinfo {year} {2007})}\BibitemShut
  {NoStop}%
\bibitem [{Sup()}]{Supplement}%
  \BibitemOpen
  \href@noop {} {}\bibinfo {note} {See the Supplemental Material for details on
  the numerical schemes, equilibrium benchmarks, and probing protocols, as well
  as additional results and discussions on the amplitude- and
  frequency-dependence of the driven correlations, linear-response checks, and
  the spatial structure of the magnetic screening currents.}\BibitemShut
  {Stop}%
\bibitem [{\citenamefont {Fertig}(2002)}]{Fertig2002Deconfinement}%
  \BibitemOpen
  \bibfield  {author} {\bibinfo {author} {\bibfnamefont {H.~A.}\ \bibnamefont
  {Fertig}},\ }\href {https://doi.org/10.1103/PhysRevLett.89.035703} {\bibfield
   {journal} {\bibinfo  {journal} {Phys. Rev. Lett.}\ }\textbf {\bibinfo
  {volume} {89}},\ \bibinfo {pages} {035703} (\bibinfo {year}
  {2002})}\BibitemShut {NoStop}%
\bibitem [{\citenamefont {Zhang}\ and\ \citenamefont
  {Fertig}(2006)}]{Zhang2006Correlation}%
  \BibitemOpen
  \bibfield  {author} {\bibinfo {author} {\bibfnamefont {W.}~\bibnamefont
  {Zhang}}\ and\ \bibinfo {author} {\bibfnamefont {H.~A.}\ \bibnamefont
  {Fertig}},\ }\href {https://arxiv.org/abs/cond-mat/0603028} {\bibinfo {title}
  {Correlation functions for the $xy$ model in a magnetic field}} (\bibinfo
  {year} {2006}),\ \Eprint {https://arxiv.org/abs/cond-mat/0603028}
  {arXiv:cond-mat/0603028 [cond-mat.other]} \BibitemShut {NoStop}%
\bibitem [{\citenamefont {Michael}\ \emph {et~al.}(2022)\citenamefont
  {Michael}, \citenamefont {F\"orst}, \citenamefont {Nicoletti}, \citenamefont
  {Haque}, \citenamefont {Zhang}, \citenamefont {Cavalleri}, \citenamefont
  {Averitt}, \citenamefont {Podolsky},\ and\ \citenamefont
  {Demler}}]{Michael2022Generalized}%
  \BibitemOpen
  \bibfield  {author} {\bibinfo {author} {\bibfnamefont {M.~H.}\ \bibnamefont
  {Michael}}, \bibinfo {author} {\bibfnamefont {M.}~\bibnamefont {F\"orst}},
  \bibinfo {author} {\bibfnamefont {D.}~\bibnamefont {Nicoletti}}, \bibinfo
  {author} {\bibfnamefont {S.~R.~U.}\ \bibnamefont {Haque}}, \bibinfo {author}
  {\bibfnamefont {Y.}~\bibnamefont {Zhang}}, \bibinfo {author} {\bibfnamefont
  {A.}~\bibnamefont {Cavalleri}}, \bibinfo {author} {\bibfnamefont {R.~D.}\
  \bibnamefont {Averitt}}, \bibinfo {author} {\bibfnamefont {D.}~\bibnamefont
  {Podolsky}},\ and\ \bibinfo {author} {\bibfnamefont {E.}~\bibnamefont
  {Demler}},\ }\href {https://doi.org/10.1103/PhysRevB.105.174301} {\bibfield
  {journal} {\bibinfo  {journal} {Phys. Rev. B}\ }\textbf {\bibinfo {volume}
  {105}},\ \bibinfo {pages} {174301} (\bibinfo {year} {2022})}\BibitemShut
  {NoStop}%
\bibitem [{\citenamefont {Mihlin}\ and\ \citenamefont
  {Auerbach}(2009)}]{Mihlin2009TemperatureDependence}%
  \BibitemOpen
  \bibfield  {author} {\bibinfo {author} {\bibfnamefont {A.}~\bibnamefont
  {Mihlin}}\ and\ \bibinfo {author} {\bibfnamefont {A.}~\bibnamefont
  {Auerbach}},\ }\href {https://doi.org/10.1103/PhysRevB.80.134521} {\bibfield
  {journal} {\bibinfo  {journal} {Phys. Rev. B}\ }\textbf {\bibinfo {volume}
  {80}},\ \bibinfo {pages} {134521} (\bibinfo {year} {2009})}\BibitemShut
  {NoStop}%
\bibitem [{Note3()}]{Note3}%
  \BibitemOpen
  \bibinfo {note} {For the correlation and optical-conductivity calculations,
  we impose periodic boundary conditions in both in-plane directions. For the
  static magnetic-field response, we instead use open boundary conditions along
  $x$ and periodic boundary conditions along $y$, as detailed in Sec.
  ``Magnetic probing protocol''.}\BibitemShut {Stop}%
\bibitem [{\citenamefont {Masini}\ \emph {et~al.}(2025)\citenamefont {Masini},
  \citenamefont {Cuccoli}, \citenamefont {Rettori}, \citenamefont
  {Trombettoni},\ and\ \citenamefont {Cinti}}]{Masini2025Helicity}%
  \BibitemOpen
  \bibfield  {author} {\bibinfo {author} {\bibfnamefont {A.}~\bibnamefont
  {Masini}}, \bibinfo {author} {\bibfnamefont {A.}~\bibnamefont {Cuccoli}},
  \bibinfo {author} {\bibfnamefont {A.}~\bibnamefont {Rettori}}, \bibinfo
  {author} {\bibfnamefont {A.}~\bibnamefont {Trombettoni}},\ and\ \bibinfo
  {author} {\bibfnamefont {F.}~\bibnamefont {Cinti}},\ }\href
  {https://doi.org/10.1103/PhysRevB.111.094415} {\bibfield  {journal} {\bibinfo
   {journal} {Physical Review B}\ }\textbf {\bibinfo {volume} {111}},\ \bibinfo
  {pages} {094415} (\bibinfo {year} {2025})}\BibitemShut {NoStop}%
\bibitem [{\citenamefont {Hasenbusch}(2005)}]{Hasenbusch2005BKT}%
  \BibitemOpen
  \bibfield  {author} {\bibinfo {author} {\bibfnamefont {M.}~\bibnamefont
  {Hasenbusch}},\ }\href {https://doi.org/10.1088/0305-4470/38/26/003}
  {\bibfield  {journal} {\bibinfo  {journal} {Journal of Physics A:
  Mathematical and General}\ }\textbf {\bibinfo {volume} {38}},\ \bibinfo
  {pages} {5869} (\bibinfo {year} {2005})}\BibitemShut {NoStop}%
\end{thebibliography}

%apsrev4-2.bst 2019-01-14 (MD) hand-edited version of apsrev4-1.bst
%Control: key (0)
%Control: author (72) initials jnrlst
%Control: editor formatted (1) identically to author
%Control: production of article title (-1) disabled
%Control: page (0) single
%Control: year (1) truncated
%Control: production of eprint (0) enabled
%

\begin{comment}
%
%
%
% End Matter
% PRL End Matter: appendixes and specialist material placed after the references.
\clearpage
\appendix
\setcounter{section}{0}
\renewcommand{\thesection}{A\arabic{section}}
\setcounter{equation}{0}
\renewcommand{\theequation}{A\arabic{equation}}
\setcounter{figure}{0}
\renewcommand{\thefigure}{A\arabic{figure}}
\setcounter{table}{0}
\renewcommand{\thetable}{A\arabic{table}}

\begin{center}
\textbf{\Large{\large{End Matter}}}
\end{center}

\textit{Appendix: ***.---}
***
\end{comment}

\clearpage
\onecolumngrid
% Supplemental Material begins here.
\appendix
\setcounter{figure}{0}
\renewcommand{\thefigure}{S\arabic{figure}}
\renewcommand{\theHfigure}{S.\arabic{figure}}
\setcounter{table}{0}
\renewcommand{\thetable}{S\arabic{table}}
\renewcommand{\theHtable}{S.\arabic{table}}
\setcounter{equation}{0}
\renewcommand{\theequation}{S\arabic{equation}}
\renewcommand{\theHequation}{S.\arabic{equation}}
\setcounter{section}{0}
\renewcommand{\thesection}{S\arabic{section}}

\begin{center}
{\large\bfseries Supplemental Material for\\``Optical and magnetic signatures of drive-enhanced
\\coherence in phase-disordered superconducting bilayers''}
\end{center}

\section{Langevin dynamics and numerical implementation}
\label{sec:supp_dynamics}

We simulate Eq.~\eqref{eq:H} on two $L\times L$ square lattices~\footnote{For the correlation 
and optical-conductivity calculations, we impose periodic boundary conditions in both in-plane directions. 
For the static magnetic-field response, we instead use open boundary conditions along $x$ and periodic 
boundary conditions along $y$, as detailed in Sec. ``Magnetic probing protocol''.}.
Here, every rescaled numerical quantity carries a superscript $*$, 
while physical quantities are written without one.  Thus
\begin{equation}
H^{*}=\frac{H}{J_0},\quad
T^{*}=\frac{k_{\rm B}T}{J_0},\quad
t^{*}=\frac{t}{\tau},\quad
J_\perp^{*}=\frac{J_\perp}{J_0},\quad
\Delta J_\ell^{*}=\frac{\Delta J_\ell}{J_0},\quad
\Omega^{*}=\Omega\tau,
\label{eq:star_defs}
\end{equation}
where $\tau=(\Gamma J_0)^{-1}$.  The dimensionless equations of motion are
\begin{align}
\partial_{t^{*}}\theta_{1i}
={}&-\bigl(1+\Delta J_1^{*}\bigr)
 \sum_{j\in i}\sin\phi_{1,ij}^{*}
 - J_\perp^{*}\sin(\theta_{1i}-\theta_{2i})
 +\xi_{1i}^{*},\nonumber\\
\partial_{t^{*}}\theta_{2i}
={}&-\bigl(1+\Delta J_2^{*}\bigr)
 \sum_{j\in i}\sin\phi_{2,ij}^{*}
 + J_\perp^{*}\sin(\theta_{1i}-\theta_{2i})
 +\xi_{2i}^{*},
\end{align}
where $\phi_{\ell,ij}^{*}=\theta_{\ell i}-\theta_{\ell j}-A_{\ell,ij}^{*}$ and
\begin{equation}
\avg{\xi_{\ell i}^{*}(t^{*})\xi_{\ell'j}^{*}(t^{*\prime})}
=2T^{*}\delta_{\ell\ell'}\delta_{ij}\delta(t^{*}-t^{*\prime}).
\end{equation}
Our implementation uses an Euler--Maruyama update,
\begin{equation}
\begin{aligned}
\theta_{\ell i}(t^{*}+\Delta t^{*})
={}&\theta_{\ell i}(t^{*})
-\Delta t^{*}\,
 \frac{\partial H^{*}}{\partial\theta_{\ell i}}\\
&+\sqrt{2T^{*}\Delta t^{*}}\,\eta_{\ell i},
\end{aligned}
\label{eq:EM}
\end{equation}
with independent unit normal variates $\eta_{\ell i}$.  Angles are wrapped to
$(-\pi,\pi]$ after each step.  We first equilibrate with
$\Delta J_1^{*}=\Delta J_2^{*}=0$, reset the production clock,
and switch on the drive with the envelope
\begin{equation}
g(t^{*})=1-\exp[-(t^{*}/t_{\rm sw}^{*})^2],
\qquad
t_{\rm sw}^{*}=n_{\rm sw}\frac{2\pi}{\Omega^{*}}.
\end{equation}
Measurements are taken only after the transient associated with the switch-on
has decayed.  The main-text data use
$\Delta t^{*}\sim0.05$,
$n_{\rm sw}\sim3$, a burn-in of $\sim10^4$ steps, and a
production window of $\sim10$ drive periods.  We verify
convergence against smaller $\Delta t^{*}$, longer burn-in and production
times, and larger $L$.

\begin{figure}[b]
\centering
\draftgraphic{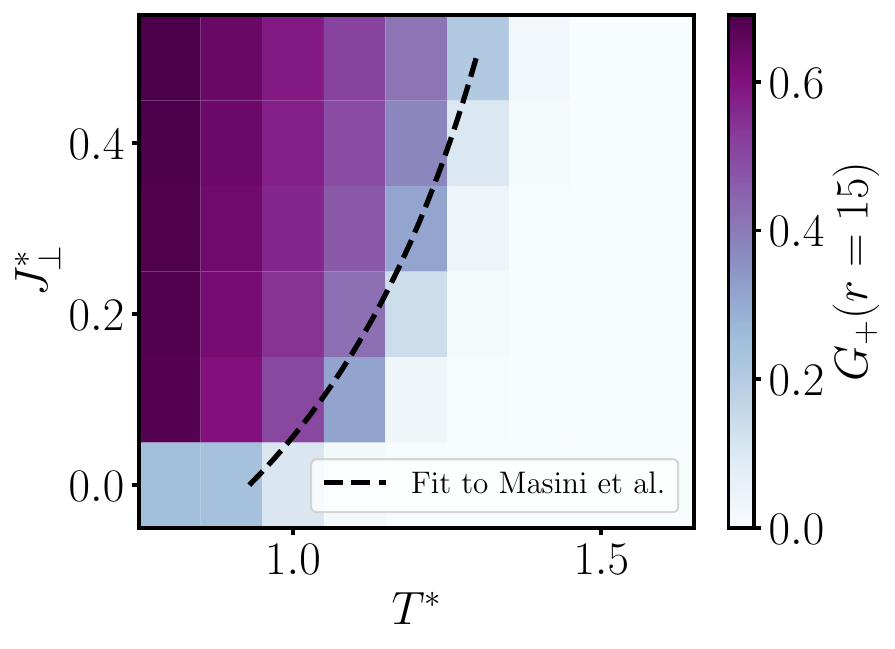}{0.35\textwidth}{Equilibrium benchmark map}
\caption{\textbf{Equilibrium benchmark in the $(T^*, J_\perp^*)$ plane.} Layer-symmetric correlations $G_+$, 
evaluated at fixed $r=15$, as a function of $T^*$ and $J_\perp^*$. At $J_\perp^{*}=0$, 
the two layers independently reproduce the square-lattice XY transition at $T_{\rm BKT}^{*}\sim0.89$. 
A finite $J_\perp^{*}$ locks the relative phase and shifts the finite-size ordering crossover 
to higher temperatures. The dashed black line represents a fit to data obtained from worm-algorithm 
calculations of the bilayer XY model by Masini et al.~\cite{Masini2025Helicity}. 
Other parameters are $L=64$ and $N=10$.}
\label{fig:supp_equilibrium_corr}
\end{figure}

The phases in Eq.~\eqref{eq:pmphase} are compact.  In the numerical
implementation we therefore use
\begin{equation}
\theta_-=\operatorname{wrap}(\theta_1-\theta_2),\qquad
\theta_+=\operatorname{wrap}\!\left[
\theta_2+\frac{1}{2}\operatorname{wrap}(\theta_1-\theta_2)\right].
\label{eq:codepm}
\end{equation}
The second expression is the midpoint of the shortest arc connecting
$\theta_1$ and $\theta_2$.  It avoids the artificial discontinuity of the
naive arithmetic average when either phase crosses the branch cut.  For 
$S_\pm(\vec{r})=e^{\ii\theta_\pm(\vec{r})}$, correlations are evaluated
under periodic boundary conditions as
$C_\pm(\vec{r})=\mathcal{F}^{-1}\!\left[|\mathcal{F}S_\pm|^2\right]/L^2$
and subsequently averaged over radial shells.

\begin{comment}
Vorticity is obtained by summing wrapped nearest-neighbor phase differences
around each plaquette. We used it as an auxiliary diagnostic of the
driven phase-ordering mechanism~\cite{DeSantis2025Enhanced}.
\end{comment}

\section{Equilibrium bilayer benchmark}
\label{sec:supp_equilibrium}

Before driving, we benchmark our framework as a function of $T^{*}$ 
and $J_\perp^{*}$. At $J_\perp^{*}=0$, the two layers independently 
reproduce the square-lattice XY transition at $T_{\rm BKT}^{*}\sim0.89$~\cite{Hasenbusch2005BKT}. 
A finite $J_\perp^{*}$ locks the relative phase and shifts the finite-size 
ordering crossover to higher temperatures. As a representative example, 
the layer-symmetric response agrees reasonably well with recent worm-algorithm 
calculations of the bilayer XY model~\cite{Masini2025Helicity}, 
as shown in Fig.~\ref{fig:supp_equilibrium_corr}.

\begin{figure}[t]
\centering
\draftgraphic{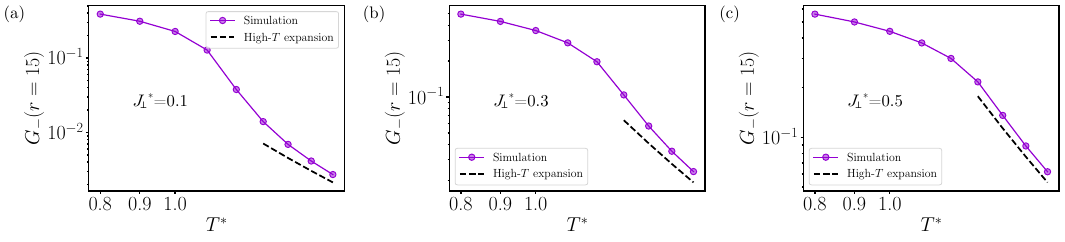}{\textwidth}{Equilibrium benchmark high-T}
\caption{\textbf{Temperature behavior of relative-phase correlations.}
Layer-antisymmetric correlations $G_-$, evaluated at fixed distance $r=15$, as a function 
of temperature $T^*$ for different interlayer couplings: (a) $J_\perp^{*}=0.1$, (b) $J_\perp^{*}=0.3$, 
and (c) $J_\perp^{*}=0.5$. Simulation results are compared with the high-temperature expansion 
derived from Refs.~\cite{Zhang2006Correlation,Michael2026Counterflow} (dashed black lines). 
Other parameters are $L=64$ and $N=10$.}
\label{fig:supp_equilibrium_highT}
\end{figure}

Regarding relative-phase locking, it is worth noting that the layer-antisymmetric 
correlation function remains finite at large distances, with a magnitude that 
depends on $J_\perp^{*}$. This behavior is illustrated in 
Fig.~\ref{fig:supp_equilibrium_highT}, where we plot $G_-$, evaluated at fixed 
distance $r=15$, as a function of temperature for different interlayer couplings: 
$J_\perp^{*}=0.1$ [panel (a)], $J_\perp^{*}=0.3$ [panel (b)], and $J_\perp^{*}=0.5$ 
[panel (c)]. The numerical results are compared with the corresponding 
high-temperature expansion derived from Refs.~\cite{Zhang2006Correlation,Michael2026Counterflow}, 
showing good agreement in the high-temperature regime.

\section{Amplitude--frequency dependence}
\label{sec:supp_phase_diagram}

\begin{figure}[t]
\centering
\draftgraphic{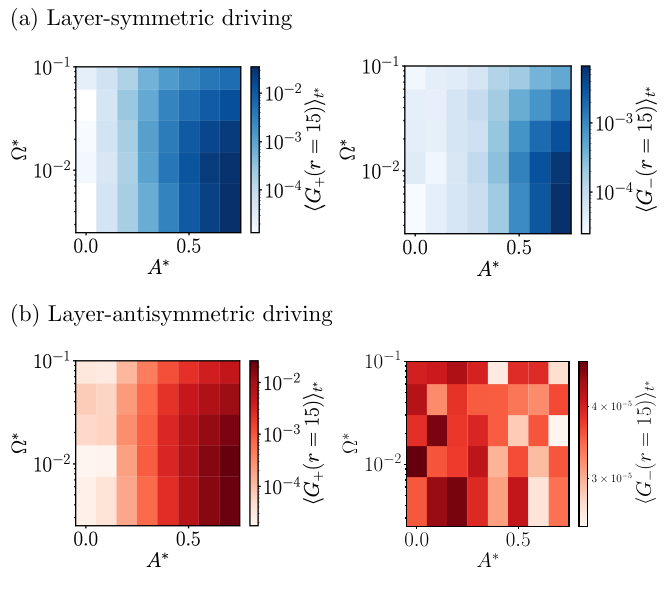}{0.55\textwidth}{Amplitude-frequency scans}
\caption{\textbf{Drive phase diagrams.} Amplitude-frequency dependence
of the time-averaged correlators $G_+(r=15)$ and $G_-(r=15)$ for
layer-symmetric [panel~(a)] and layer-antisymmetric modulation
[panel~(b)]. Slow, strong driving enhances $G_+$ for both drive
symmetries, whereas enhancement of $G_-$ is confined to the
layer-symmetric drive. Other parameters are $L=64$, $N=10$,
$T^*=1.5$, and $J_\perp^*=0.01$.}
\label{fig:supp_phase_diagram}
\end{figure}

We scan $A^{*}$ and $\Omega^{*}$ at fixed $T^{*}$ and
$J_\perp^{*}$, and compute the time-averaged correlators at large
separation. The results are shown in Fig.~\ref{fig:supp_phase_diagram}
for layer-symmetric [panel~(a)] and layer-antisymmetric modulation
[panel~(b)].

For $G_+$, the enhancement is largest in the strongly driven,
quasi-adiabatic regime for both drive symmetries. When $\Omega^{*}$
exceeds the collective phase-relaxation scale, the system can no longer
develop significant correlations over large distances. This frequency
dependence reproduces the single-layer mechanism~\cite{DeSantis2025Enhanced}.
Thus, slow and strong driving provides the most favorable regime for
enhancing $G_+$, independently of the drive symmetry.

The behavior of $G_-$ is different. A clear enhancement
is observed only under layer-symmetric driving, again favoring the
strongly driven, quasi-adiabatic regime. Under layer-antisymmetric
driving, by contrast, the large-distance $G_-$ correlations show no
enhancement and no systematic structure in amplitude-frequency
parameter space.

\section{Optical probing protocol}
\label{sec:supp_conductivity}

%\subsection{Physical probe and dimensionless implementation}
For a physical $x$-polarized probe,
\begin{equation}
A_p(t)=\frac{2eaE_0}{\hbar\omega}\cos(\omega t+\phi),
\qquad
E_p(t)=E_0\sin(\omega t+\phi),
\end{equation}
where $A_p$ is the Peierls phase generated by the physical electric field
$E_p$.  The corresponding dimensionless simulation variables are
\begin{equation}
t^{*}=t/\tau,\qquad \omega^{*}=\omega\tau,\qquad
E_p^{*}(t^{*})=\frac{2ea\tau}{\hbar}E_p(t),
\end{equation}
so that
\begin{equation}
A_p^{*}(t^{*})=\frac{E_0^{*}}{\omega^{*}}
\cos(\omega^{*}t^{*}+\phi),\qquad
E_p^{*}(t^{*})=E_0^{*}\sin(\omega^{*}t^{*}+\phi),
\end{equation}
with $E_0^{*}=2eaE_0\tau/\hbar$ and
$E_p^{*}=-\partial_{t^{*}}A_p^{*}$.  At each measurement time we accumulate
\begin{equation}
\begin{aligned}
j_+^{*}(\omega^{*})
&=\Delta t_{\rm m}^{*}\sum_n e^{\ii\omega^{*}t_n^{*}}j_+^{*}(t_n^{*}),\\
E_+^{*}(\omega^{*})
&=\Delta t_{\rm m}^{*}\sum_n e^{\ii\omega^{*}t_n^{*}}E_p^{*}(t_n^{*}),
\end{aligned}
\end{equation}
and form the dimensionless conductivity
\begin{equation}
\sigma^{*}(\omega^{*})=
\frac{j_+^{*}(\omega^{*})}{E_+^{*}(\omega^{*})}.
\end{equation}
A separate stochastic trajectory is generated for every probe frequency. The
probe is weak compared with the drive and is applied only after equilibration.

\begin{figure}[t]
\centering
\draftgraphic{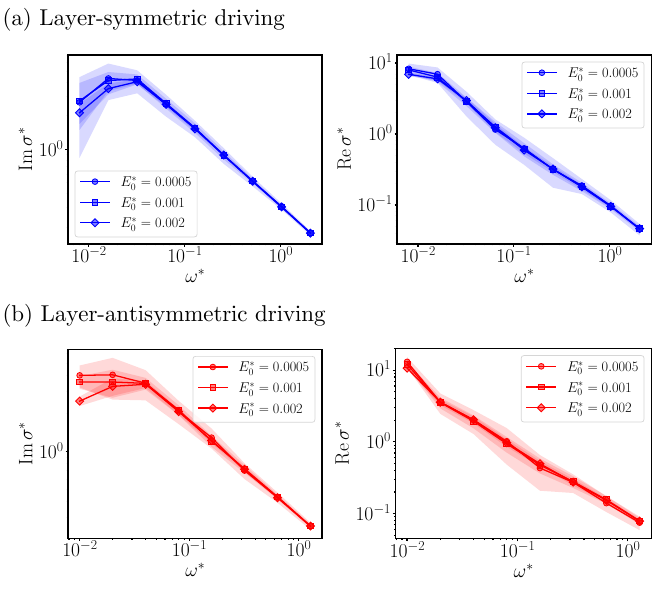}{0.55\textwidth}{Conductivity linearity}
\caption{\textbf{Linearity of the driven optical response.}
Imaginary and real parts of the conductivity $\sigma^*$ under layer-symmetric 
[panel~(a)] and layer-antisymmetric driving [panel~(b)], 
evaluated under the favorable conditions identified in the main text. 
Results are shown for probe amplitudes $E_0^*=0.0005$ (circles), $0.001$ (squares), and $0.002$ (diamonds). 
The extracted conductivity remains essentially unchanged as the probe amplitude 
is varied, confirming that the system is probed within the linear-response regime. 
Shaded areas denote the statistical error over realizations. 
Other parameters are $L=64$, $N=50$, $T^*=1.5$, $J_\perp^*=0.01$, $\Omega^*=0.01$, and $A^*=0.7$.}
\label{fig:supp_cond_lin}
\end{figure}

To verify that the optical response is evaluated within the linear-response regime, 
we systematically vary the amplitude of the weak probe field while keeping the driving protocol 
unchanged. Figure~\ref{fig:supp_cond_lin} compares the conductivity obtained,
under representative driving conditions, for probe amplitudes $E_0^*=0.0005$ (circles), 
$0.001$ (squares), and $0.002$ (diamonds), for both layer-symmetric and 
layer-antisymmetric driving. Despite the fourfold variation in probe amplitude, 
the extracted real and imaginary parts of the conductivity remain essentially 
unchanged. This confirms that the results discussed in the main text are 
insensitive to the probe strength and correspond to the linear optical 
response of the driven system.

\begin{comment}
\subsection{Conversion to physical conductivity}
The physical sheet current and electric field are related to their
simulation counterparts by
\begin{equation}
K_\ell(t)=\frac{2eJ_0}{\hbar a}\,j_\ell^{*}(t^{*}),
\qquad
E_p(t)=\frac{\hbar}{2ea\tau}\,E_p^{*}(t^{*}).
\end{equation}
Consequently,
\begin{equation}
\sigma_{2\mathrm D}(\omega)
=\sigma_0\,\sigma^{*}(\omega^{*}),
\qquad
\sigma_0=\left(\frac{2e}{\hbar}\right)^2J_0\tau.
\label{eq:sigma_conversion}
\end{equation}
For comparison with a three-dimensional optical conductivity, one should divide by
an effective layer spacing,
\begin{equation}
\sigma_{3\mathrm D}(\omega)=
\frac{\sigma_{2\mathrm D}(\omega)}{d_{\rm eff}}.
\label{eq:sigma3Dunits}
\end{equation}
\end{comment}

\section{Magnetic probing protocol}
\label{sec:supp_screening}

\begin{figure}[b]
\centering
\draftgraphic{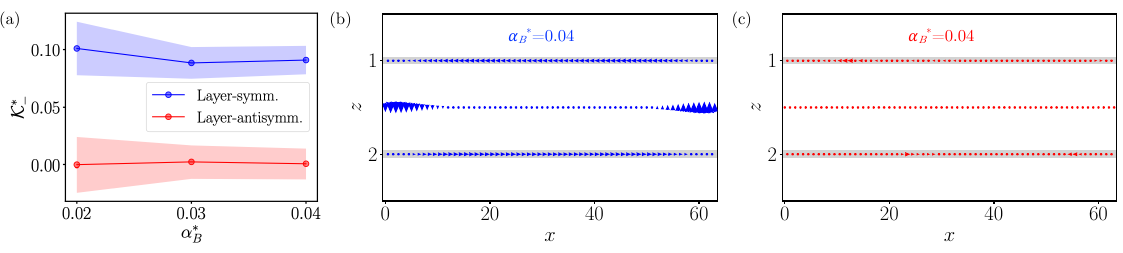}{\textwidth}{Screening linearity and current profiles}
\caption{\textbf{Linear-response regime and spatial structure of the magnetic screening.}
(a) Screening kernel $\mathcal K_-^*$ for three strengths of the applied
vector potential, $\alpha_B^*=0.02$, $0.03$, and $0.04$, under
layer-symmetric (blue) and layer-antisymmetric (red) driving. The weak
dependence on $\alpha_B^*$ confirms that the response lies within the
linear regime over the range considered. (b) Time- and realization-averaged
current profile in the $x$--$z$ plane for layer-symmetric driving at
$\alpha_B^*=0.04$, showing the drive-induced screening loop.
(c) Corresponding profile for layer-antisymmetric driving, for which no
appreciable screening loop develops. Arrow lengths in panels (b) and (c)
are normalized independently for visualization. Shaded areas in panel (a)
indicate the statistical uncertainty over realizations. Other parameters
are $L=64$, $N=10$, $T^*=1.5$, $J_\perp^*=0.02$,
$\Omega^*=0.005$, and $A^*=0.8$.}
\label{fig:supp_screen_lin}
\end{figure}

\begin{figure}[t]
\centering
\draftgraphic{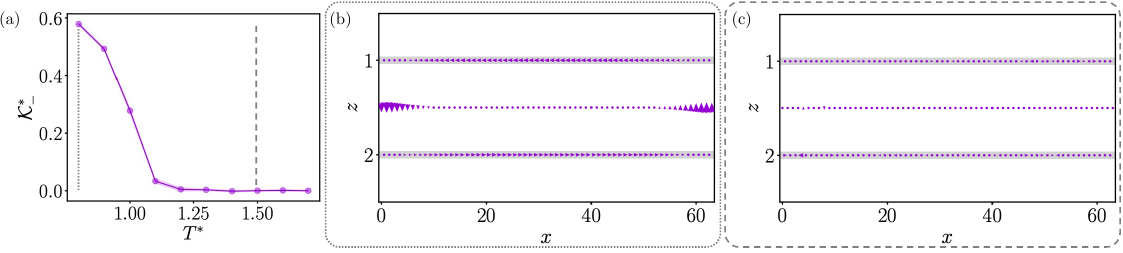}{\textwidth}{Equilibrium screening benchmark}
\caption{\textbf{Equilibrium benchmark of the magnetic screening.}
(a) Equilibrium screening kernel $\mathcal K_-^*$ as a function of
temperature. The response decreases rapidly across the ordering crossover
and becomes negligible in the high-temperature regime. (b) Time- and
realization-averaged current profile in the $x$--$z$ plane at $T^*=0.8$,
showing the closed screening-current pattern below the transition.
(c) Corresponding profile at $T^*=1.5$, where no appreciable screening
loop remains. Arrow lengths in panels (b) and (c) are normalized
independently for visualization. Shaded areas in panel (a) indicate the
statistical uncertainty over realizations. Other parameters are $L=64$,
$N=10$, $J_\perp^*=0.02$, and $\alpha_B^*=0.04$.}
\label{fig:supp_screen_eq}
\end{figure}

For the physical field $\mathbf B=B_y\mathbf e_y$, we choose
$A_x(z)=B_yz$ and place the layers at $z=\pm d_\perp/2$. The corresponding
dimensionless Peierls phases are
\begin{equation}
A_{1x}^{*}=-A_{2x}^{*}=\alpha_B^{*},
\qquad
\alpha_B^{*}=\frac{ead_\perp B_y}{\hbar}.
\end{equation}
For the magnetic-response calculations, we impose open boundary conditions
along the $x$ direction, parallel to the applied vector potential, and
periodic boundary conditions along $y$. The absence of bonds connecting
the two $x$ edges allows the screening currents to form explicitly in the
finite sample and, for finite $J_\perp^*$, to close through interlayer
Josephson currents near the boundaries.

The static magnetic perturbation is present throughout the magnetic
simulation. In particular, the system is first evolved through a burn-in
stage in the presence of $A_{1x}^{*}=-A_{2x}^{*}=\alpha_B^*$, allowing the
phase configuration and screening currents to relax in the applied field.
After this field-on burn-in, we reset the production clock and either
monitor the equilibrium magnetic response or switch on the periodic
stiffness modulation. Thus, the driven results characterize the response
of an initially field-relaxed state rather than the transient associated
with suddenly applying the magnetic perturbation.

The dimensionless layer currents along $x$ are evaluated on the physical
nearest-neighbor bonds as
\begin{equation}
j_{\ell x}^{*}(x,y,t^*)
=
J_\ell^{*}(t^*)
\sin\!\left[
\theta_\ell(x,y,t^*)-\theta_\ell(x+1,y,t^*)
-A_{\ell x}^{*}
\right],
\end{equation}
and their spatially averaged counterflow combination is
\begin{equation}
j_-^{*}=\frac{j_1^{*}-j_2^{*}}{2}.
\end{equation}
The screening kernel used in the main text is then
\begin{equation}
\mathcal K_-^{*}
=-\lim_{\alpha_B^{*}\to0}
\frac{\overline{\avg{j_-^{*}}}}{\alpha_B^{*}},
\label{eq:supp_Kminus}
\end{equation}
where the overbar denotes the production-time average and
$\avg{\cdots}$ includes the spatial and realization averages.

The open geometry also allows us to resolve the spatial structure of the
screening response directly. We define the interlayer Josephson current
at each in-plane site by
\begin{equation}
j_z^{*}(x,y,t^*)
=
J_\perp^{*}
\sin\!\left[
\theta_1(x,y,t^*)-\theta_2(x,y,t^*)
\right],
\end{equation}
with positive $j_z^*$ defined as flowing from layer 1 to layer 2. For the
current-loop plots in Figs.~\ref{fig:supp_screen_lin} and
\ref{fig:supp_screen_eq}, the currents are first averaged along the
periodic $y$ direction, over the production window, and over independent
realizations. Horizontal arrows on layers 1 and 2 represent the resulting
$j_{1x}^*(x)$ and $j_{2x}^*(x)$ on the corresponding $x$ bonds, while
vertical arrows represent $j_z^*(x)$ on the interlayer bonds. Since layer
1 is drawn above layer 2, positive $j_z^*$ points downward in the plots.
For clarity, the arrows in each panel are normalized by the largest
absolute current in that panel; their directions and relative lengths
therefore display the spatial circulation pattern, whereas absolute
screening strengths should be compared through $\mathcal K_-^*$.

The same magnetic perturbation can alternatively be represented in a gauge
with $A_x=0$ and $A_z=-B_yx$. In physical units, the interlayer term then
takes the form
\begin{equation}
-J_\perp\sum_i\cos\!\left[
\theta_{1i}-\theta_{2i}
-\frac{2e}{\hbar}d_\perp B_yx_i
\right].
\end{equation}
The magnetic field therefore imposes a long-wavelength twist of the
relative phase $\theta_-$, making explicit that the magnetic response is
controlled by relative-phase coherence. In the open geometry this same
response appears directly as oppositely directed in-plane currents in the
two layers connected by interlayer Josephson currents, forming a closed
screening loop.

We verify the linear-response regime by repeating the driven calculation
for several strengths of the applied vector potential. As shown in
Fig.~\ref{fig:supp_screen_lin}(a), $\mathcal K_-^*$ is essentially
unchanged for $\alpha_B^*=0.02$, $0.03$, and $0.04$. The magnetic response
reported in the main text can therefore be identified with the linear
screening response. The spatial profiles provide a complementary view:
under layer-symmetric driving, Fig.~\ref{fig:supp_screen_lin}(b) displays
a clear circulating current pattern, with counterpropagating in-plane
currents connected by interlayer currents near the open boundaries.
Under layer-antisymmetric driving, by contrast,
Fig.~\ref{fig:supp_screen_lin}(c) shows no appreciable circulation,
consistent with the near-vanishing $\mathcal K_-^*$ in panel~(a).

As an equilibrium benchmark, we evaluate the undriven magnetic response
as a function of temperature. Figure~\ref{fig:supp_screen_eq}(a) shows
that $\mathcal K_-^*$ decreases rapidly on crossing the equilibrium
ordering regime and becomes very small at high temperature. The
corresponding spatial profiles make this loss of screening particularly
transparent. At $T^*=0.8$, below the transition,
Fig.~\ref{fig:supp_screen_eq}(b) exhibits a well-developed screening loop,
whereas at $T^*=1.5$ the circulation is absent within our numerical
resolution [Fig.~\ref{fig:supp_screen_eq}(c)]. The latter temperature is
the equilibrium starting point used for the driven calculations in the
main text. The screening loop generated there by layer-symmetric driving
therefore represents a genuine drive-induced enhancement relative to the
initial high-temperature state.

\begin{comment}
\subsection{Conversion to magnetic susceptibility}
The physical counterflow sheet current is
\begin{equation}
K_{\rm cf}=\frac{2eJ_0}{\hbar a}\,j_-^{*},
\end{equation}
and the physical linear-response coefficient is
$\mathcal K_-=-K_{\rm cf}/B_y$ in the $B_y\to0$ limit. Two opposite sheet
currents separated by $d_\perp$ carry a magnetic moment per unit area
$m_y/\mathcal A=K_{\rm cf}d_\perp$. Dividing by the bilayer repeat distance
$d_{\rm bilayer}$ gives
\begin{equation}
\chi_y\equiv\frac{M_y}{H_y}
\simeq\mu_0\frac{M_y}{B_y}
=-\mu_0\frac{d_\perp}{d_{\rm bilayer}}\,\mathcal K_-.
\label{eq:chiunits}
\end{equation}
\end{comment}

\end{document}